\documentclass[prd,onecolumn,english]{revtex4}
\usepackage{babel}

\usepackage{amsmath}
\usepackage{amssymb}
\usepackage{amsthm} % Для оформления теорем, задач и прочих окружений.

\usepackage{array}

\usepackage{slashed}

\usepackage[dvips]{graphicx} % Поддержка вставки графических файлов.
\usepackage{subfigure}

\usepackage{varioref}   % Расширенные ссылки со страницами.
\usepackage{xr}         % Ссылки на внешние документы.

\usepackage{pstricks}           % PSTricks with the `color' interface

\allowdisplaybreaks[1]

\newcommand\nn{\nonumber}
\newcommand\ba{\begin{eqnarray}}
\newcommand\ea{\end{eqnarray}}
\newcommand\eq[1] {\begin{align} #1 \end{align}}   % Уравнение без рамочки.
\newcommand\ga[1] {\begin{gather} #1 \end{gather}}   % Уравнение без рамочки.

\newcommand{\br}[1]{\left( #1 \right)}
\newcommand{\brs}[1]{\left[ #1 \right]}
\newcommand{\brf}[1]{\left\{ #1 \right\}}
\newcommand{\brm}[1]{\left| #1 \right|}

\newcommand{\Sp}{\mbox{Sp}}

\newcommand{\vv}[1]{{\bf #1}}

\newcommand{\dd}[1]{{\hat #1}}   % Свертка с gamma-матрицей Дирака.

\newcommand{\GeV}{\mbox{GeV}}

\newcommand{\g}[1] {{#1}}

\newcommand{\Z}[1] {{\bf #1}}

\newcommand{\M} {{\cal M}} % Матричный элемент.

\renewcommand{\a} {{\mathbf a}} % Знаменатели петлевых интегралов.
\renewcommand{\b} {{\mathbf b}} % Знаменатели петлевых интегралов.
\renewcommand{\c} {{\mathbf c}} % Знаменатели петлевых интегралов.

\newcommand{\Moller} {{M{\o}ller}~} % Мёллер.
\begin{document}
% =========================================================================

\title{Parity violating \Moller scattering asymmetry up to the two-loop level}

\author{A.~G. Aleksejevs}
\affiliation{Memorial University, Corner Brook, Canada}
\email{aaleksejevs@grenfell.mun.ca}

\author{S.~G. Barkanova}
\affiliation{Acadia University, Wolfville, Canada}
\email{svetlana.barkanova@acadiau.ca}

\author{Yu. M.~Bystritskiy}
\affiliation{Joint Institute for Nuclear Research, Dubna, Russia}
\email{bystr@theor.jinr.ru}

\author{A. N.~Ilyichev}
\affiliation{National Center of Particle and High Energy Physics of Belarussian State University, Minsk, Belarus}
\email{ily@hep.by}

\author{E. A.~Kuraev}
\affiliation{Joint Institute for Nuclear Research, Dubna, Russia}
\email{kuraev@theor.jinr.ru}

\author{V. A.~Zykunov}
\affiliation{Belarusian State University of Transport, Gomel, Belarus}
\email{vladimir.zykunov@cern.ch}

\date{\today}

\begin{abstract}
The paper investigates contributions of $Z$, $W$ and $\gamma$ intermediate states
to the parity-violating \Moller scattering asymmetry  up to two-loop level.
Using the Yennie--Frautschi--Suura factorization form for amplitudes,
we demonstrate that QED corrections, with an exception of vacuum-polarization
effects, cancel at the asymmetry level.
We obtain chiral amplitudes at Born, one-loop and partially at two-loop level:
boxes with lepton self-energies, ladder boxes and decorated boxes.
Our calculations are relevant for the ultra-precise 11 GeV MOLLER
experiment planned at Jefferson Laboratory and future ILC experiments.
The numerical comparision of the two-loop contributions with the experimental accuracy of MOLLER is provided.
\end{abstract}

\maketitle

% =========================================================================
\section{Introduction}
\label{SectionIntroduction}
% =========================================================================

The Standard Model of Particle Physics (SM) introduces a non-zero asymmetry between
left- and right-handed particles and predicts a parity-violating (PV) interference
between the electromagnetic and weak neutral current amplitudes.
By measuring this  small asymmetry, precision experiments with polarized electron beams
(polarized M{\o}ller scattering)
attract especially active interest from both experimental and theoretical communities
as they can provide indirect access to physics at multi-TeV scales and
play an important complementary role to the LHC research program.

The MOLLER (Measurement Of a Lepton Lepton Electroweak Reaction) experiment planned
at Jefferson Lab aims to measure the parity-violating asymmetry in the scattering of
$11~\GeV$ longitudinally-polarized electrons from the atomic electrons in a liquid
hydrogen target (\Moller scattering) with a combined statistical and systematic
uncertainty of 2\%~\cite{vanOers:2010zz, MOLLER2011, JLab12, Kumar:2009zzk}.
At such precision, any inconsistency with the SM predictions will clearly
signal new physics. However, a comprehensive analysis of radiative corrections
is needed before any conclusions can be made. Since MOLLER's stated precision
goal is significantly more ambitious than that of its predecessor E-158 \cite{Kumar:1995ym, Kumar:2007zze, Anthony:2003ub},
theoretical input for this measurement must include not only a full treatment of one-loop
(next-to-leading order, NLO) electroweak radiative corrections but also leading %% dominant? The use of 'leading' seems a bit ambiguous here.
two-loop corrections
(next-to-next-leading order, NNLO).
A significant theoretical effort has been dedicated to the one-loop radiative corrections already
\cite{Mo:1968cg, Maximon:1969nw, Czarnecki:1995fw, Denner:1998um, Petriello:2002wk, Erler:2004in, Aleksejevs:2010ub, Aleksejevs:2010nf}
(the squares of the one-loop diagrams were calculated in \cite{Aleksejevs:2011de}),
but more needs to be done on the two-loop corrections.

The main goal of this paper is to verify the previous
theoretical predictions for the \Moller asymmetry using the Yennie--Frautschi--Suura
factorization technique, and to obtain an estimation of some leading two-loop corrections:
boxes with lepton self-energies, ladder (double) boxes, and decorated boxes.
We show that at the next-to-leading order, the main contribution to the \Moller asymmetry comes from the
process with $Z$ and $W$ gauge bosons in the intermediate state and most of the pure-QED
contributions cancel out. Using the hypothesis of factorization of soft and hard contributions,
similar to that of the Drell--Yan parton picture, we calculate the SM electroweak corrections at
the NLO and partially at the next-to-NLO levels.
Detailed and consistent consideration of all two-loop corrections will be the next task of our group:
these are the combined self-energy (SE) and vertex contributions, double SEs, decorated and double vertices,
and boxes with vertex and SE insertions.
Our calculations are performed using a one-mass shell renormalization scheme in the t'Hooft--Feynman gauge.

The paper is organized as follows. In Section II, we state the PV \Moller scattering asymmetry
in the Born approximation. Section III briefly discusses approximations we use in calculating
the electroweak radiative corrections. Section IV outlines our treatment of infrared-divergent
contributions. The Yennie--Frautschi--Suura (YFS) irreducible diagrams are evaluated in Section V.
One-loop and some two-loop radiative corrections are given in Sections VI and VII, correspondingly.
The results are gathered and analyzed in Section VIII, and are followed by conclusions in Section IX.
Details and examples of our calculations are given in the Appendices.

% =========================================================================
\section{Asymmetry in the Born approximation}
\label{SectionBorn}
% =========================================================================

The \Moller process first studied by \cite{Moeller:1932} is the process of
electron--electron scattering defined as follows:
\eq{
    e^-\br{p_1,\lambda_1} + e^-\br{p_2,\lambda_2} \to e^-\br{p_1',\lambda_1'} + e^-\br{p_2',\lambda_2'}.
    \label{eq:MollerScattering}
}
The amplitude (matrix element) of this process within the SM has four terms
(see Fig.~\ref{Fig:BornApproximation}):
\begin{figure}
    \centering
    \mbox{
        \subfigure[]{\includegraphics[width=0.2\textwidth]{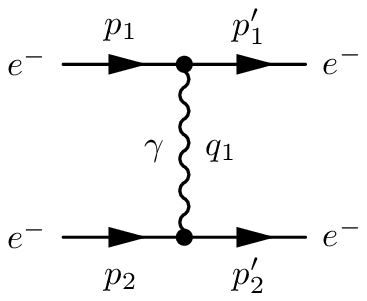}\label{FigBornGamma1}}
        \quad
        \subfigure[]{\includegraphics[width=0.2\textwidth]{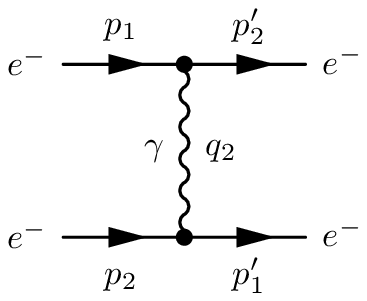}\label{FigBornGamma2}}
        \quad
        \subfigure[]{\includegraphics[width=0.2\textwidth]{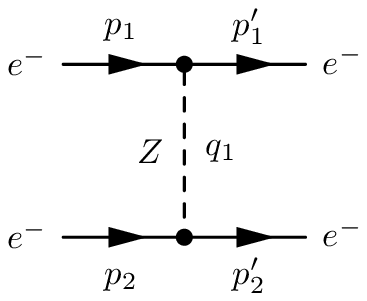}\label{FigBornZ1}}
        \quad
        \subfigure[]{\includegraphics[width=0.2\textwidth]{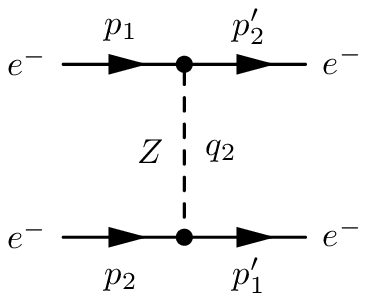}\label{FigBornZ2}}
    }
    \caption{Born approximation diagrams.}
    \label{Fig:BornApproximation}
\end{figure}
\eq{
    \M^{(0)} = \M^{(0)}_{\gamma_1} - \M^{(0)}_{\gamma_2} + \M^{(0)}_{Z_1} - \M^{(0)}_{Z_2},
}
where $M^{(0)}_{\gamma_1}$ is the contribution from a photon exchange in the $t$-channel
(Fig.~\ref{FigBornGamma1}) and
$M^{(0)}_{\gamma_2}$ is the contribution from a photon exhange in the $u$-channel, where the
two final leptons are interchanged ($p_1' \leftrightarrow p_2'$) to accommodate the Pauli principle (see Fig.~\ref{FigBornGamma2}).
The last two terms, $M^{(0)}_{Z_(1,2)}$ are similar contributions with a $Z$ boson exchange in the $t$- and $u$-channels (see Fig.~\ref{FigBornZ1},\ref{FigBornZ2}).
Using the t'Hooft--Feynman gauge, i.e. $\xi=1$, we get the following form for these terms:

\eq{
    \M^{(0)}_{\gamma_1} &= \frac{e^2}{q_1^2}
    \brs{\bar u^{\br{\lambda_1'}}\br{p_1'} \gamma^\mu u^{\br{\lambda_1}}\br{p_1}}\brs{\bar u^{\br{\lambda_2'}}\br{p_2'} \gamma_\mu u^{\br{\lambda_2}}\br{p_2}}, \nn\\
    \M^{(0)}_{\gamma_2} &= \frac{e^2}{q_2^2}
    \brs{\bar u^{\br{\lambda_2'}}\br{p_2'} \gamma^\mu u^{\br{\lambda_1}}\br{p_1}}\brs{\bar u^{\br{\lambda_1'}}\br{p_1'} \gamma_\mu u^{\br{\lambda_2}}\br{p_2}}, \nn\\
    \M^{(0)}_{Z_1} &= \frac{g^2}{16 \cos^2\theta_W}
    \brs{\bar u^{\br{\lambda_1'}}\br{p_1'} \gamma_\mu \br{a - \gamma_5} u^{\br{\lambda_1}}\br{p_1}}\brs{\bar u^{\br{\lambda_2'}}\br{p_2'} \gamma^\mu \br{a_V - \gamma_5} u^{\br{\lambda_2}}\br{p_2}}
    \frac{1}{q_1^2 - M_Z^2}, \nn\\
    \M^{(0)}_{Z_2} &= \frac{g^2}{16 \cos^2\theta_W}
    \brs{\bar u^{\br{\lambda_2'}}\br{p_2'} \gamma_\mu \br{a - \gamma_5} u^{\br{\lambda_1}}\br{p_1}}\brs{\bar u^{\br{\lambda_1'}}\br{p_1'} \gamma^\mu \br{a_V - \gamma_5} u^{\br{\lambda_2}}\br{p_2}}
    \frac{1}{q_2^2 - M_Z^2}, \nn
}
where $q_1 = p_1 - p_1' = p_2' - p_2$ and $q_2 = p_1 - p_2' = p_1' - p_2$ are the transferred momenta in the direct and exchanged diagrams,
$a_V = 1 - 4 \sin^2 \theta_W$,  $\theta_W$ is the Weinberg angle ($g \sin\theta_W = e$), and $e$ is the electric charge of a positron.

In the chiral amplitude approach we are using, the specific chiral
spin states of the initial and final particles are selected as
\eq{
    u^{\br{\lambda}} = \omega_\lambda u,
    \qquad
    \bar u^{\br{\lambda}} = \bar u \, \omega_{-\lambda},
    \qquad
    \lambda = \pm 1 = R,L,
}
where chirality projection operators $\omega_\lambda$ have the following form:
\eq{
    \omega_\lambda = \frac{1}{2}\br{1+\lambda\gamma_5},
    \qquad
    \omega_\lambda^2 = \omega_\lambda,
    \qquad
    \omega_+ \omega_- = 0,
    \qquad
    \omega_+ + \omega_- = 1.
}
For the case of massless fermions, we need to satisfy the completeness condition:
\eq{
    u^{\br{\lambda}}\br{p}\bar{u}^{\br{\lambda}}\br{p}=\omega_\lambda \dd{p}.
}
Let us calculate the QED contribution of a right+right  $\to$  right+right ($ ++ \to ++ $) chiral amplitude,
where all fermions are right-handed, i.e. $u^R=\omega_+u$:
\eq{
    \M^{(0)++++}_{\gamma_1} &= \frac{e^2}{t}
    \brs{\bar u\br{p_1'} \omega_- \gamma^\mu \omega _+ u\br{p_1}}\brs{\bar u\br{p_2'} \omega_- \gamma_\mu \omega_+ u\br{p_2}}, \nn
    \\
    \M^{(0)++++}_{\gamma_2} &= \frac{e^2}{u}
    \brs{\bar u\br{p_2'} \omega_- \gamma^\mu \omega _+ u\br{p_1}}\brs{\bar u\br{p_1'} \omega_- \gamma_\mu \omega_+ u\br{p_2}}. \nn
}
Here, we have used the Mandelstam invariants in the limit of vanishing electron mass ($m \to 0$):
\eq{
    s = 2 \br{p_1 p_2} = 2 \br{p_1' p_2'}, \quad
    t = -2 \br{p_1 p_1'} = -2 \br{p_2 p_2'}, \quad
    u = -2 \br{p_1 p_2'} = -2 \br{p_2 p_1'}, \quad
    s+t+u = 0.
}
In order to transform these amplitudes into calculable traces,
we multiply terms which contain a factor ${1/t}$ by the following quantity:
\eq{
    \frac{a \, b}{a \, b}= 1, \label{EqFactorAB}
}
and the terms with factor ${1/u}$ by
\eq{
    \frac{c \, d}{c \, d}= 1,\label{EqFactorCD}
}
where
\eq{
    a &= \bar u\br{p_1} \omega_- \dd{p_2} \omega_+ u\br{p_2'},  &c &= \bar u\br{p_1} \omega_- \dd{p_2} \omega_+ u\br{p_1'}, \label{EqABCD}\\
    b &= \bar u\br{p_2} \omega_- \dd{p_1} \omega_+ u\br{p_1'},  &d &= \bar u\br{p_2} \omega_- \dd{p_1} \omega_+ u\br{p_2'}. \nn
}
This allows us to obtain the trace in the numerator and calculate it immediately:
\eq{
    \M^{(0)++++}_{\gamma_1} &= \frac{e^2}{t}
    \frac{1}{a\, b}
    \Sp\brs{\dd{p_1'}\gamma_\mu \omega_+ \dd{p_1} \dd{p_2} \omega_+ \dd{p_2'} \gamma^\mu \omega_+ \dd{p_2} \dd{p_1} \omega_+}
    =
    \frac{e^2}{t}
    \frac{1}{a \, b}
    2 s^2 t, \\
    \M^{(0)++++}_{\gamma_2} &= \frac{e^2}{u}
    \frac{1}{c\, d}
    \Sp\brs{\dd{p_2'}\gamma_\mu \omega_+ \dd{p_1} \dd{p_2} \omega_+ \dd{p_1'} \gamma^\mu \omega_+ \dd{p_2} \dd{p_1} \omega_+}
    =
    \frac{e^2}{u}
    \frac{1}{c \, d}
    2 s^2 u. \nn
}
Thus, the QED amplitude in the Born approximation has the form
\eq{
    \M^{(0)++++}_{\gamma} = \M^{(0)++++}_{\gamma_1} - \M^{(0)++++}_{\gamma_2} =
    2\br{4\pi\alpha} s^2 \M^0_\gamma,
    \qquad
    \M^0_\gamma = \frac{1}{a \, b} - \frac{1}{c \, d}.
    \label{EqAmpBornGamma}
}
The rest of the spiral amplitudes are calculated in a similar way and lead to the following results:
\eq{
    \M^{(0)+-+-}_{\gamma} &= 2(4\pi\alpha)\frac{u^2}{t c_1d_1}, \\
    \M^{(0)+--+}_{\gamma} &= -2(4\pi\alpha)\frac{t^2}{u c_1d_1}, \nn
}
where $c_1$ and $d_1$ are the modified factors similar to (\ref{EqFactorAB}) or (\ref{EqFactorCD}), which in this case have the form:
\eq{
    c_1 &= \bar u\br{p_1} \omega_- \br{p_2'},  &d_1 &= \bar u\br{p_2} \omega_+ u\br{p_1'}. \label{EqC1D1}
}
Using relations
\eq{
|a|^2=|b|^2=-st, \qquad |c|^2=|d|^2=-su, \qquad a b c^* d^* = -s^2 t u,
}
we obtain the well-known result \cite{Baier:1973,Akhiezer:1981} for the sum of squares of all six amplitudes:
\eq{
    \sum_{\br{\lambda}}\brm{\M_{\gamma}^{\br{0}\lambda}}^2
    &=
    8\br{4\pi\alpha}^2
    \brs{
        \br{\frac{s^2}{t^2}+\frac{s^2}{u^2}+\frac{2s^2}{tu}}+
        \br{\frac{t^2}{u^2}+\frac{u^2}{t^2}}
    }=
    8\br{4\pi\alpha}^2\frac{s^4+t^4+u^4}{t^2u^2}.
}
Employing the same procedure for the amplitudes with a mediating Z boson, we obtain
\eq{
    \M^{(0)++++}_{Z}
    = \M^{(0)++++}_{Z_1} - \M^{(0)++++}_{Z_2}
    = -\frac{2 s^2}{M_Z^2} \frac{4\pi\alpha}{4 \sin^2\br{2\theta_W}} \br{1+a_V}^2 \M^0_Z,
    \qquad
    \M^0_Z = \frac{t}{a \, b} - \frac{u}{c \, d}.
    \label{EqAmpBornZ}
}
The expressions for the amplitude $\M^{----}_{Z}$ can be obtained from (\ref{EqAmpBornGamma}) and (\ref{EqAmpBornZ})
by replacing the factor $(a_V+1)^2$ with $(a_V-1)^2$.

The high-precision \Moller scattering experiments allow a careful study of the SM predictions by measuring
the polarization asymmetry defined in the standard way as
\eq{
A \equiv A_{LR} =
 \frac{\sigma_{LL}+\sigma_{LR}-\sigma_{RL}-\sigma_{RR}}
      {\sigma_{LL}+\sigma_{LR}+\sigma_{RL}+\sigma_{RR}}
 =
 \frac{\sigma_{LL}-\sigma_{RR}}
      {\sigma_{00}},
\label{A}
}
where $\sigma$ means the differential cross section ($\sigma \equiv d\sigma/dc$),  $c=\cos\br{\widehat{\vv{p_1},\vv{p_1'}}}$,
and the index $00$ corresponds to unpolarized scattering.
Using the language of chiral amplitudes, this asymmetry reads as
\eq{
    A =
    \frac{\sigma^{----} - \sigma^{++++}}
         {\sigma^{++++} + \sigma^{+--+} + \sigma^{+-+-}  + \sigma^{-++-} + \sigma^{-+-+}  + \sigma^{----}}.
}
In the Born approximation, the only contribution to this asymmetry comes from an
interference between $\M^{(0)}_\gamma$ and $\M^{(0)}_Z$, which is proportional to
\eq{
    \br{\M^0_\gamma}^*\M^0_Z
    =
    \br{\frac{1}{a \, b} - \frac{1}{c \, d}}^*\br{\frac{t}{a \, b} - \frac{u}{c \, d}}
    =
    -\frac{2}{s t u}.
}
This gives us the following expression for the Born asymmetry:
\eq{
    A^{(0)} =
    \frac{s}{2 M_W^2}
    A_0 \frac{1-4\sin^2\theta_W}{\sin^2 \theta_W},
    \qquad
    A_0=
    \frac{y\br{1-y}}{1+y^4+\br{1-y}^4}
    ,
    \qquad
    y =\frac{-t}{s}=\frac{1-c}{2}.
    \label{AsymmetryInBorn}
}
In spite of this asymmetry being extremely small ($\sim 10^{-7}$),
the accuracy of modern and upcoming experiments clearly exceeds the accuracy of the
theoretical result in the Born approximation. In addition,
one--loop contributions to the the parity-violating
\Moller scattering asymmetry were found to be very large
\cite{Denner:1998um, Czarnecki:1995fw, Aleksejevs:2010ub},
which points to the extreme importance of the careful inclusion of the higher-order radiative corrections.

% =========================================================================
\section{General discussion of radiative corrections}
\label{SectionGeneralDiscussion}
% =========================================================================

Radiative corrections are the higher-order contributions to the leading-order Feynman diagrams,
and their inclusion is an essential part of any modern experiment. In this work, we consider the
SM and QED radiative corrections at the one- and two-loop levels.
Although some progress \cite{Aleksejevs:2010ub, Aleksejevs:2011de} has recently been achieved in calculating radiative
corrections for \Moller scattering with semi-automated computer-algebra packages like FeynArts \cite{Hahn:2000kx},
FormCalc \cite{Hahn:1998yk}, LoopTools \cite{Hahn:1998yk} and Form \cite{Vermaseren:2000nd}, we believe that since work on the two-loop
corrections is still at an early stage, it is prudent to do careful and explicit derivations first, with
semi-automated results to follow later.

Obviously, at this stage some approximations are unavoidable. A helpful approximation we employ throughout
this work is based on the effective factorization of contributions from the emission of real soft photons and
virtual photons with small virtuality \cite{Yennie:1961ad}. In this approximation, we can omit all Feynman
diagrams with virtual photons connecting external (on-mass-shell) electron lines. The same statement is valid
for the emission of real soft photons. This approximation significantly reduces the number of diagrams we have to
evaluate. The relevant modification of the Born asymmetry $\tilde A$ is discussed in
Section~\ref{SectionHardSubdiagramsCalculation}. Another simplification we use is neglecting the dependence on
the external momenta.

In addition to bremsstrahlung, the types of  Feynman amplitudes we consider at one-loop level are vacuum
polarization diagrams (Fig.~\ref{Fig:OneLoopPolarization}), boxes (Fig.~\ref{Fig:OneLoopBox}), and vertex
corrections (Fig.~\ref{Fig:OneLoopVertexCorrection}).
At the two-loop level, we evaluate the self-energy insertions into lepton lines in the two-boson exchange
boxes and the two types of the double-box diagrams -- ladder type and decorated-box type.
The following contributions from SM corrections are to be considered later:
the boxes with 1) vertices and 2) SE insertions,
3) combined SE and vertex contributions, 4) double SEs, 5) decorated and 6) double vertices.

% =========================================================================
\section{Extraction of the infrared-divergent part}
\label{SectionInfraredPartExtraction}
% =========================================================================

A comprehensive and detailed analysis of infrared-divergent contributions for a general
case was performed by \cite{Yennie:1961ad}. Following the \cite{Yennie:1961ad} findings,
we express contributions from infrared-divergent radiative corrections in the form of an
exponent convoluted with the infrared-finite hard subprocess part of the amplitude.

\begin{figure}
    \centering
    \mbox{
        \subfigure[]{\includegraphics[width=0.16\textwidth]{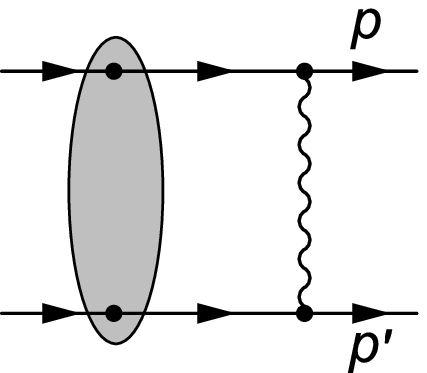}\label{FigISRVirt}}
        \quad
        \subfigure[]{\includegraphics[width=0.16\textwidth]{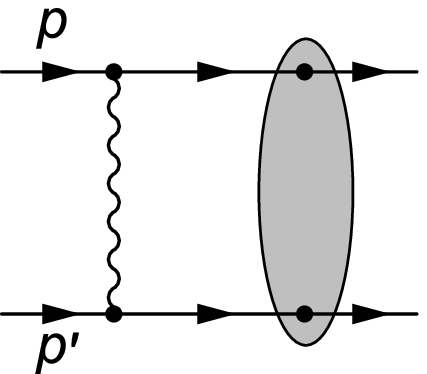}\label{FigFSRVirt}}
        \quad
        \subfigure[]{\includegraphics[width=0.16\textwidth]{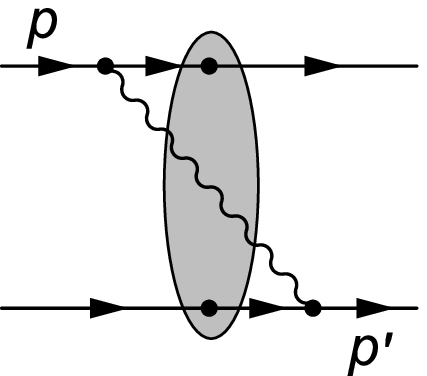}\label{FiguChannelCorrection}}
        \quad
        \subfigure[]{\includegraphics[width=0.16\textwidth]{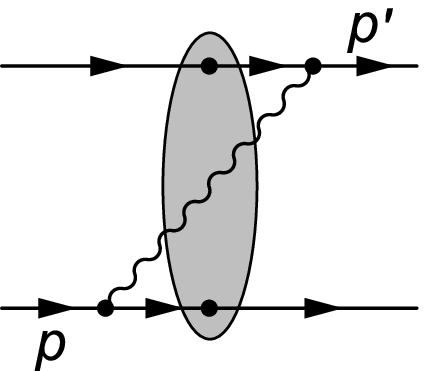}\label{FiguChannelCorrection2}}
    }\\
    \mbox{
        \subfigure[]{\includegraphics[width=0.16\textwidth]{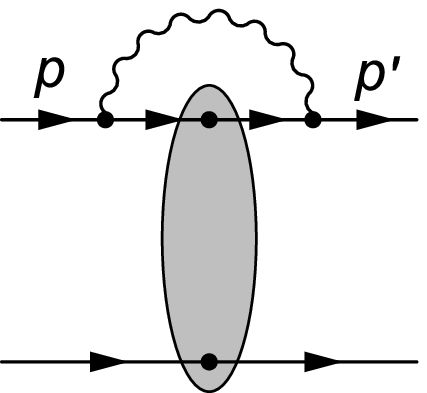}\label{FigtChannelCorrection}}
        \quad
        \subfigure[]{\includegraphics[width=0.16\textwidth]{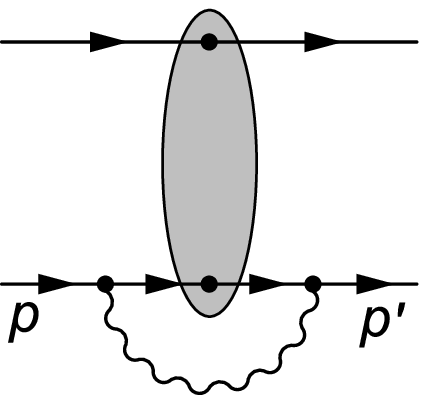}\label{FigtChannelCorrection2}}
    }
    \caption{Demonstration of YFS irreducible hard subprocess in first order of perturbation theory
            (see for details Section~\ref{SectionInfraredPartExtraction}).}
    \label{Fig:YFSIrreducibleDiagrams}
\end{figure}

Let us review some of the \cite{Yennie:1961ad} results which we use in this work.
The amplitude of any process with external (ingoing and outgoing)
charged particles has the form:
\eq{
    \M\br{p,p'} = \sum_{n=0}^\infty \M_n,
}
where $p$ and $p'$ are the external on-mass shell particle momenta
(see for, example, Fig.~\ref{Fig:YFSIrreducibleDiagrams} notation), a summation is done over different
orders of perturbation theory contributions coming from the emission of $n$ virtual photons, and
$\M_n$ is the amplitude of the process in the
$n$-th order of perturbation (i.e. proportional to $e^n$, where $e$ is the positron charge).

It has been proven that the amplitudes $\M_n$ have the following structure:
\eq{
    \M_0 &= m_0, \nn\\
    \M_1 &= \alpha B \, m_0 + m_1, \nn\\
    \M_2 &= \frac{\br{\alpha B}^2}{2!} m_0 + \alpha B \, m_1 + m_2, \nn\\
    &\cdots\nn\\
    \M_n &= \sum_{r=0}^n \frac{\br{\alpha B}^r}{r!} m_{n-r}, \label{eq:AmplitudeMn}
}
where $m_0$ is the amplitude in the Born approximation and $m_n$ ($n>0$) are the
infrared-finite pieces of the amplitude of $n$-th order in the perturbation theory
(we call it \emph{hard subprocess amplitude}). Fig.~\ref{Fig:YFSIrreducibleDiagrams}
illustrates the amplitude $m_1$ with the hard subprocess showed by
filled blocks.
The term $B$ introduced in (\ref{eq:AmplitudeMn}) has the form:
\eq{
    B = \frac{i}{\br{2\pi}^3} \int \frac{d^4 k}{k^2-\lambda^2}
    \br{
        \frac{2p'_\mu-k_\mu}{2\br{p' k}-k^2}
        -
        \frac{2p_\mu-k_\mu}{2\br{p k}-k^2}
    }^2,
}
where $k$ and $\lambda$ are the momentum and the photon mass parameter, which we will take to the zero limit later.
Thus using the structure of the $n$-th order amplitude defined by (\ref{eq:AmplitudeMn}), for the total
amplitude we can write:
\eq{
    \M =
    \sum_{n=0}^\infty \sum_{r=0}^n \frac{\br{\alpha B}^r}{r!} m_{n-r}
    =
    \sum_{r=0}^\infty \frac{\br{\alpha B}^r}{r!} \sum_{n=0}^\infty  m_{n}
    =
    \exp\br{\alpha B} \sum_{n=0}^\infty  m_{n}.
}
The cross section of the process then reads as
\eq{
    \sigma = \exp\br{2\alpha\br{B+\tilde B}} \hat\sigma = \exp\br{\delta_t} \hat\sigma, \label{CSexp}
}
where $\hat\sigma$ is the \emph{hard process} cross section which is finite in the limit $\lambda \to 0$.
The additional term in (\ref{CSexp}) $\tilde B$ takes into
account the infrared-divergent part of the real soft-photon emission:
\eq{
    \tilde B = -\frac{1}{8\pi^2} \int_{}^{'} \frac{d^3 {\bf k}}{{ \sqrt{{\bf k}^2+\lambda^2} }}
    \br{
        \frac{p'_\mu}{\br{p' k}}
        -
        \frac{p_\mu}{\br{p k}}
    }^2,
}
where the accent at the integral denotes the region $|\bf{k}| < \omega$.
The parameter  $\omega $ is the maximum energy of real photons which escape undetected; it
is defined by a specific experimental setup.

The sum of the contributions from the virtual and real soft-photon emission in (\ref{CSexp}) is
\eq{
    \delta_t
    \equiv
    2\alpha ({B+\tilde B}) =
    -\frac{2\alpha}{\pi}
    \br{l_t-1} \ln{\frac{\sqrt{s}}{2\omega}}
    +
    \frac{\alpha}{2\pi} l_t,
    \qquad
    l_t = \ln{\frac{-t}{m^2}},
    \label{eq:deltat}
}
This sum is infrared-stable, i.e. finite in the $\lambda \to 0$ limit, which is a manifestation of
the well-known cancelation requirement of infrared singularities described by \cite{Bloch:1937pw}.

For \Moller scattering (\ref{eq:MollerScattering}), the soft-photon emission
factor can be transformed in the following manner:
\eq{
    \br{
        -\frac{{p_1  }^{\mu}}{\br{p_1 k}}
        +\frac{{p_1' }^{\mu}}{\br{p_1' k}}
        -\frac{{p_2  }^{\mu}}{\br{p_2 k}}
        +\frac{{p_2' }^{\mu}}{\br{p_2' k}}
    }^2
    &=
    \br{
        \frac{ {p_1  }^{\mu}}{\br{p_1 k}}
        -\frac{{p_1' }^{\mu}}{\br{p_1' k}}
    }^2
    +
    \br{
        \frac{ {p_2  }^{\mu}}{\br{p_2 k}}
        -\frac{{p_2' }^{\mu}}{\br{p_2' k}}
    }^2 -
    \nn\\
    &-
    \br{
        \frac{ {p_1 }^{\mu}}{\br{p_1 k}}
        -\frac{{p_2 }^{\mu}}{\br{p_2 k}}
    }^2
    -
    \br{
        \frac{ {p_1' }^{\mu}}{\br{p_1' k}}
        -\frac{{p_2' }^{\mu}}{\br{p_2' k}}
    }^2 +
    \nn\\
    &+
    \br{
        \frac{ {p_1 }^{\mu}}{\br{p_1 k}}
        -\frac{{p_2'}^{\mu}}{\br{p_2' k}}
    }^2
    +
    \br{
        \frac{ {p_2 }^{\mu}}{\br{p_2 k}}
        -\frac{{p_1'}^{\mu}}{\br{p_1' k}}
    }^2.
}
Combining (\ref{CSexp}) and (\ref{eq:deltat}), we can now write out the infrared cancellation in
the cross section in the form:
\eq{
    \sigma = \exp\br{2 \brs{ \delta_t+\delta_u-\delta_s } } \hat \sigma ,
    \label{Eq:YFSCrossSection}
}
where $\delta_{u,s}$ can be obtained from $\delta_t$ (see (\ref{eq:deltat}))
by replacing $l_t \to l_{u,s}$ with $l_u=\ln\br{-u/m^2}$ and $l_s=\ln\br{s/m^2}$.
First term in the exponent (with $\delta_t$) gives the contribution of the diagrams from
Fig.~\ref{FigtChannelCorrection} and \ref{FigtChannelCorrection2},
while $\delta_u$ represents diagrams from Fig.~\ref{FiguChannelCorrection} and \ref{FiguChannelCorrection2}.
Third term (with $\delta_s$) corresponds to the diagrams from Fig.~\ref{FigISRVirt} and \ref{FigFSRVirt}.
Expanding this result on $\alpha/\pi$, we get expressions identical to our previous formulas
for the first order ((55) from \cite{Aleksejevs:2010ub})
\eq{
\frac{\sigma^{\rm NLO}}{\sigma^0} =
\frac{2\alpha}{\pi}  \ln \frac{4\omega^2}{s} \br{\ln\frac{tu}{m^2s} -1 }+ ...
}
and the Q-part of the second order ((45) from \cite{Aleksejevs:2011de})
%Should Q-part be defined, even with the reference?
%
\eq{
\frac{\sigma_{Q}}{\sigma^0} =
\frac{1}{2} \frac{\sigma^{\rm NNLO}+ ...}{\sigma^0} =
\frac{1}{2} \br{\frac{\alpha}{\pi}}^2
\brs{ 2 \ln\frac{4\omega^2}{s} \br{ \ln\frac{tu}{m^2s} -1 } }^2+ ...\ .
}
To separate the "soft" and "hard" types of higher-order contributions, we use the factorized form of the
cross section shown in (\ref{Eq:YFSCrossSection}). The
"hard" contribution is essentially determined by the presence of the SM heavy bosons with large momentum $q$ in loops;
$m^2 \ll \brm{q^2} \sim M_Z^2$, so we simplify calculations by neglecting the dependence on the external momenta.

% =========================================================================
\section{Calculation of hard subdiagrams}
\label{SectionHardSubdiagramsCalculation}
% =========================================================================

As it was shown in Section~\ref{SectionInfraredPartExtraction}, the infrared-divergent terms are extracted
from the amplitude as an exponential factor.
We will refer to such contributions as {\it factorized}, and
for them
\begin{eqnarray}
\sigma_{ij}^{f}= \delta_f \sigma_{ij}^{0},\ ij = LL, RR, LR, 00.
\label{fact}
\end{eqnarray}
If the physical contribution $C$ to the observable asymmetry $A$
is determined by the relative correction to the Born asymmetry,
\begin{eqnarray}
\delta^{\rm C}_A = (A^{\rm C}-A^{(0)})/A^{(0)},
\label{del-a}
\end{eqnarray}
then it is clear that these factorized contributions do not change the asymmetry, as
they cancel each other in a ratio of a nominator to a denominator: $\delta^{f}_A=0$.
The contributions without the factorized property (\ref{fact})
will be referred to as {\it non-factorized}.
The real physical cross section is the sum denoted by index $f+n$
\begin{eqnarray}
\sigma_{ij}^{f+n}= \sigma_{ij}^{f} + \sigma_{ij}^{n}.
\end{eqnarray}

We should mention that in the general case  $\delta^{f+n}_A \not = \delta^{f}_A + \delta^{n}_A = \delta^{n}_A$,
but for the correct sum it is necessary to use the following formula:
\begin{eqnarray}
\delta^{f+n}_A
= \delta^{n}_A \frac{\sigma_{00}^0+\sigma_{00}^n}{\sigma_{00}^0+\sigma_{00}^f+\sigma_{00}^n}
%= \delta^{n}_A \frac{1+\delta_n}{1+\delta_f+\delta_n},
= \delta^{n}_A \Bigl( 1 - \frac{\delta_f}{1+\delta_f+\delta_n} \Bigr),
\label{non-a}
\end{eqnarray}
where $\delta_n = \sigma_{00}^{n}/\sigma_{00}^{0}$.
It is the key formula we used for taking into consideration the radiative corrections up to the two-loop level.
{A recipe is very simple}:
 the cross sections coming from the two-loop contributions are small, so
1) the  relative corrections $\delta_f$ and $\delta_n$ from (\ref{non-a}) are determined by the one-loop corrections, only
2) we add the two-loop contributions as {\it additive } terms
(we used this terminology of our work \cite{Aleksejevs:2010nf})
of relative correction $\delta^{n}_A$
to the one-loop corrections obtained under firm control before (see, for example, our paper \cite{Aleksejevs:2010ub}).
The contributions which should be evaluated explicitly
are the hard subprocess terms (terms $m_i$ from (\ref{eq:AmplitudeMn})).
The contribution which comes from a photon emitted from an inner part of this hard diagram does not contain any
infrared-divergent parts.
As it is was proven in \cite{Yennie:1961ad}, all infrared-divergent contributions come from diagrams
with a virtual photon connecting outer legs of charged particles (see, for example,
Fig.~\ref{FigBoxGG1} -- \ref{FigBoxZG2}, Fig.~\ref{FigPropagatorGammaVertexG} and Fig.~\ref{FigPropagatorZVertexG}).
Thus we only need to calculate the \emph{Yennie--Frautschi--Suura---irreducible} (\emph{YFS--irreducible}) diagrams as
they do not contain contributions with a virtual photon connecting two external electron lines.

Another useful approximation we employ is including only hard virtual photons in
$\hat\sigma$. This is similar to the factorization of soft and hard virtual corrections in
the Drell--Yan cross section done in \cite{Lipatov:1974qm,Altarelli:1977zs,Kuraev:1985hb}.

Except for the ultraviolet-divergent subdiagrams of vertex, lepton self-energy and boson vacuum
polarization insertions, the skeleton contribution to $m_i$ from (\ref{eq:AmplitudeMn})
is ultraviolet-convergent. A regularization scheme must be applied to the
ultraviolet-divergent subdiagrams; we use the on-shell renormalization scheme and the
t'Hooft--Feynman gauge.
Also, we assume that the loop momenta relevant to the skeleton amplitudes are large in comparison with
the external momenta
$\brm{\chi^2} \gg s \sim -t \sim -u \gg m^2$, so we neglect the external momenta in $m_i$.

% =========================================================================
\section{One-loop radiative corrections}
\label{SectionOneLoopRC}
% =========================================================================

The one-loop cross section was evaluated carefully and with full, firm control
of the uncertainties in the literature; see, for example, our recent paper \cite{Aleksejevs:2010ub}.
Here, we present some of the one-loop results which are relevant to this work; in particular, we want to compare the chiral amplitude method
and the approach suggested in \cite{Aleksejevs:2010ub, Aleksejevs:2010nf, Aleksejevs:2011de}.

% =========================================================================
%\subsection{Self energy diagrams}
%\label{SubSectionSE}
% =========================================================================

\begin{figure}
    \centering
    \mbox{
        \subfigure[]{\includegraphics[width=0.15\textwidth]{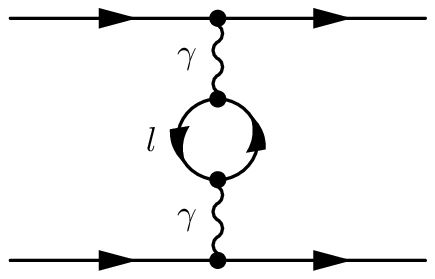}\label{FigPropagatorGammaVPL}}
        \quad
        \subfigure[]{\includegraphics[width=0.15\textwidth]{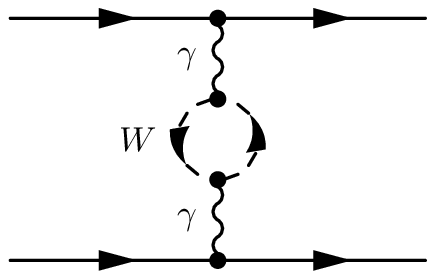}\label{FigPropagatorGammaVPW}}
        \quad
        \subfigure[]{\includegraphics[width=0.15\textwidth]{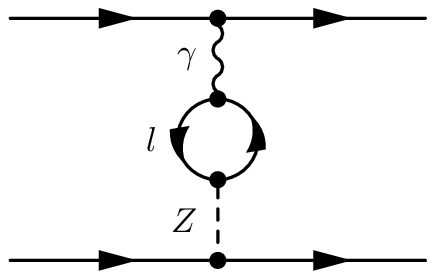}\label{FigPropagatorGammaZVPL}}
        \quad
        \subfigure[]{\includegraphics[width=0.15\textwidth]{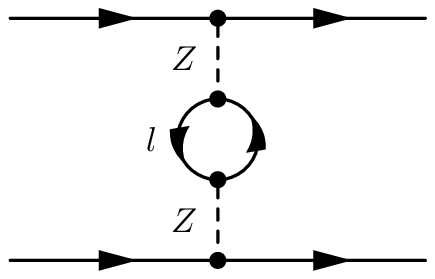}\label{FigPropagatorZVPL}}
        \quad
        \subfigure[]{\includegraphics[width=0.15\textwidth]{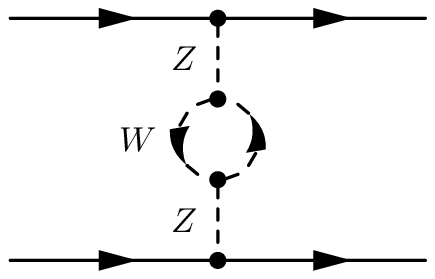}\label{FigPropagatorZVPW}}
        }
    \caption{One-loop vacuum polarization diagrams.}
    \label{Fig:OneLoopPolarization}
\end{figure}

Vacuum polarizations of virtual photons which must be taken into account in $m_i$
are shown by Fig.~\ref{Fig:OneLoopPolarization}(a) and Fig.~\ref{Fig:OneLoopPolarization}(b).
For the bilinear combination of amplitudes entering the asymmetry in the Born approximation,
this leads to the following replacement (see \ref{AsymmetryInBorn}):
\eq{
    A_{\gamma SE} = A^{(0)}|_{A_0 \rightarrow \bar A_0},
}
where
\eq{
    \bar A_0 = \frac{y(1-y)(A_t(1-y)+A_uy)}{(A_t(1-y)+A_uy)^2+A_u^2y^4+A_t^2(1-y)^4}.
    \label{AsymVPModification}
}
The self-energy (SE) factors associated with the $t$- and $u$-channels of the amplitudes
in the Born approximation, $A_t$ and $A_u$ , have the form \cite{Akhiezer:1981}:
\eq{
A_t&=\frac{1}{1-\Pi_t}, \qquad A_u=\frac{1}{1-\Pi_u}, \nn\\
\Pi_t&=\frac{\alpha}{3\pi}\br{l_t-\frac{5}{3}}+
\frac{\alpha^2}{4\pi^2}\br{l_t+\zeta_3-\frac{5}{24}} + \cdots, \qquad l_t=\ln\frac{-t}{m^2}.\nn\\
\Pi_u&= \Pi_t\br{l_t\to l_u}, \nn
}
where $\zeta_3 \approx 1.202$ is the Riemann zeta function.
The asymmetry that includes the Born approximation as well as the first- and the second-order corrections can be written as:
$    {A} =
    A_{\gamma SE} \bigl( 1 + \frac{\alpha}{\pi} \delta_1 +
    \br{\frac{\alpha}{\pi}}^2 \delta_2 \bigr)$.
The term $\delta_1$ contains contributions from the one-loop diagrams with two-$Z$,
$WZ$, and $WW$ exchange, as well as the vertex functions of leptons with the
$Z$ and $W$ bosons in the intermediate state.
The term $\delta_2$ contains contributions from the self-energy insertions
into the lepton functions with $W$ and $Z$ boson exchanges, boson vacuum polarizations and
two-loop Feynman diagrams of two types -- double-box and
decorated-box.

% =========================================================================
%\subsection{Self-energy-type diagrams}
%\label{SubSectionSelfEnTypeDiagrams}
% =========================================================================

The contribution to the vacuum polarization from the $W$-boson in the intermediate state
$A_\gamma^\Pi$ (Fig. 3 (b) and (e))
does not add anything to the asymmetry as it has the same form for the
$\br{----}$ and $\br{++++}$ spiral states:
\eq{
    A_\gamma^\Pi &=
    0.\nn
}

% =========================================================================
%\subsection{Box-type diagrams}
%\label{SubSectionBoxTypeDiagrams}
% =========================================================================

\begin{figure}
    \centering
    \mbox{
        \subfigure[]{\includegraphics[width=0.15\textwidth]{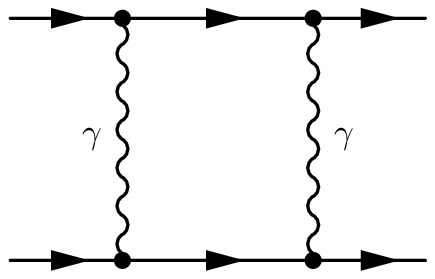}\label{FigBoxGG1}}
        \quad
        \subfigure[]{\includegraphics[width=0.15\textwidth]{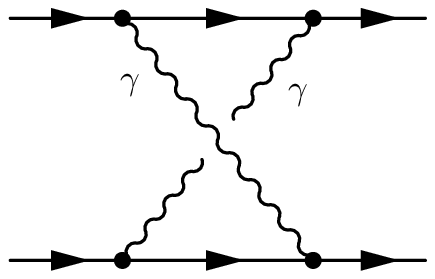}\label{FigBoxGG2}}
        \quad
        \subfigure[]{\includegraphics[width=0.15\textwidth]{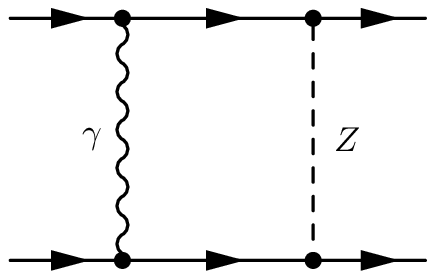}\label{FigBoxGZ1}}
        \quad
        \subfigure[]{\includegraphics[width=0.15\textwidth]{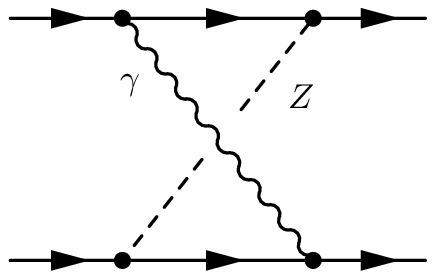}\label{FigBoxGZ2}}
        \quad
        \subfigure[]{\includegraphics[width=0.15\textwidth]{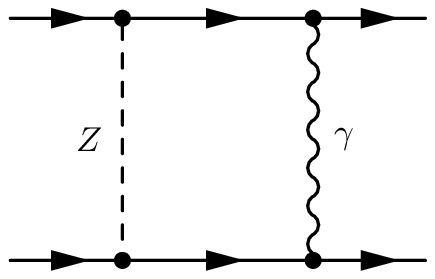}\label{FigBoxZG1}}
    }
    \\
    \mbox{
        \subfigure[]{\includegraphics[width=0.15\textwidth]{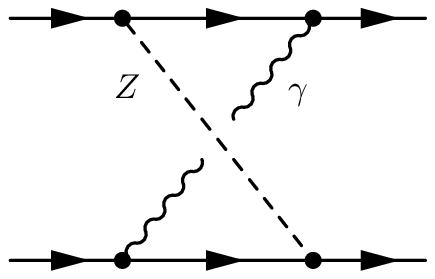}\label{FigBoxZG2}}
        \quad
        \subfigure[]{\includegraphics[width=0.15\textwidth]{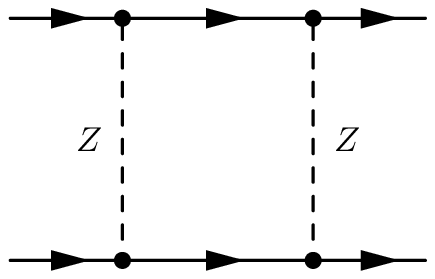}\label{FigBoxZZ1}}
        \quad
        \subfigure[]{\includegraphics[width=0.15\textwidth]{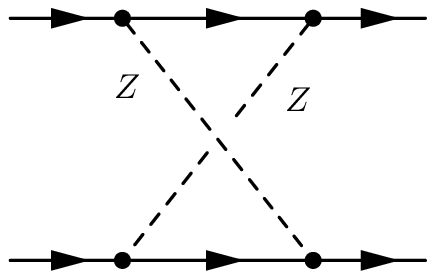}\label{FigBoxZZ2}}
        \quad
        \subfigure[]{\includegraphics[width=0.15\textwidth]{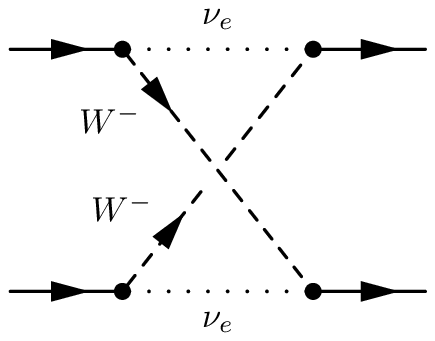}\label{FigBoxWW}}
        }
    \caption{One-loop box type diagrams.}
    \label{Fig:OneLoopBox}
\end{figure}

With our approach, we only need to consider the YFS--irreducible diagrams,
i.e. we can omit the contributions from the YFS--reducible diagrams
Fig.~\ref{FigBoxGG1} -- \ref{FigBoxZG2}.
Thus it is sufficient to consider only one diagram with crossed $W$-legs,
shown in Fig.~\ref{FigBoxWW}, which contributes to the $\M^{----}$ amplitude. The other two, Fig.~\ref{FigBoxZZ1} and Fig.~\ref{FigBoxZZ2},
 lead to the second term in $\M_Z^0$ from (\ref{EqAmpBornZ}).
Let us start with the $WW$ crossed box:
\eq{
    2\M_B^* \M_{WW}
    =
    4s^4
    \br{-4\alpha\pi i} \frac{\br{4\pi\alpha}^2 i\pi^2}{4s_W^4 \br{2\pi}^4}
    \br{\M_\gamma^0}^* \M_Z^0 N_{WW} \frac{1}{M_Z^2},
}
where $s_W = \sin\theta_W$, $c_W = \cos\theta_W$ and
\eq{
    N_{WW} &= \frac{M_W^2}{{s^2 t}}  \int d\chi
    \frac{S_{WW}}{\br{\chi^2}^2\br{\chi^2-M_W^2}^2},
    \qquad
    d\chi = \frac{d^4 \chi}{i\pi^2}, \label{NWW}
    \\
    S_{WW} &=
    \Sp\brs{
        \dd{p_1'}\gamma_\mu \dd{\chi} \gamma_\nu
        \dd{p_1}\dd{p_2} \dd{p_2'} \gamma^\nu \dd{\chi} \gamma^\mu \dd{p_2} \dd{p_1}
        \omega_-
    }
    =
    2 \chi^2 s^2 t.\nn
}
We use a Wick rotation to perform the loop momenta integration in (\ref{NWW}):
\eq{
    d\chi \to \chi_e^2 d\chi_e^2,
    \qquad
    \text{where}
    \qquad
    \chi^2 = -\chi_e^2 < 0.
}
Evaluating the integral, we obtain   $ N_{WW} = -2$.
Thus, the resultant contribution to the asymmetry is:
\eq{
    A_{WW}
    =
    \frac{\alpha}{\pi} \bar A_0
    \frac{s}{8M_W^2}
     \frac{1}{s_W^4}.
}
This result is in full agreement with the relative correction to the asymmetry
induced by a $WW$-box obtained in \cite{Aleksejevs:2010nf} (see formula (40) there):
$A_{WW}=\delta^{WW}_A A^{(0)}$,
where $\delta^{C}_A$ is determined by (\ref{del-a}).
For all {\it non-factorized} two-loop corrections,
the formula connecting the relative correction $\delta^{C}_A$
with the asymmetry induced by the effect $C$ is
\begin{equation}
A_{C}=\delta^{C}_A A^{(0)}.
\end{equation}

A contribution from the $ZZ$ box and crossed $ZZ$ box diagrams,
(Fig.~\ref{FigBoxZZ1} and \ref{FigBoxZZ2}), has a similar form:
\eq{
    A_{ZZ}
    =
    -
    \frac{\alpha}{\pi} \bar A_0
    \frac{s}{M_Z^2}
     \frac{a_V }{32 c_W^4 s_W^4} N_{ZZ},
    \qquad
    N_{ZZ} = 6.
}
Similarly to the $WW$-box case, the relative correction to the asymmetry
induced by a $ZZ$-box obtained in \cite{Aleksejevs:2010nf} is in perfect agreement
with the result presented here:
$A_{ZZ}=\delta^{ZZ}_A A^{(0)}$.

% =========================================================================
%\subsection{Vertex-type diagrams}
%\label{SubSectionVertexTypeDiagrams}
% =========================================================================

\begin{figure}
    \centering
    \mbox{
        \subfigure[]{\includegraphics[width=0.15\textwidth]{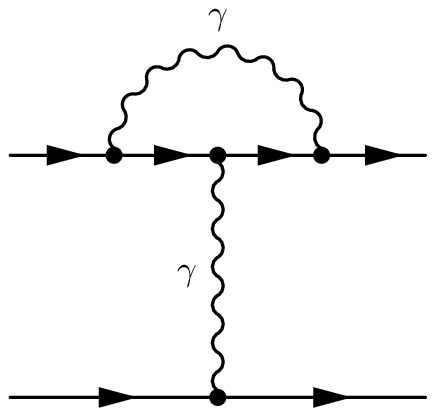}\label{FigPropagatorGammaVertexG}}
        \quad
        \subfigure[]{\includegraphics[width=0.15\textwidth]{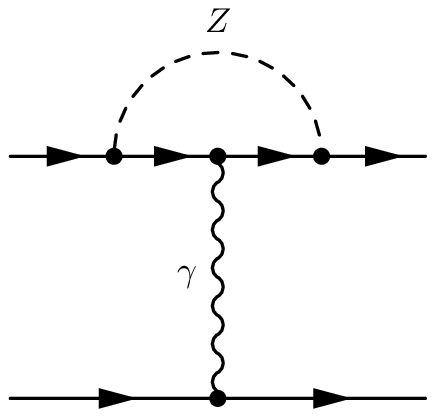}\label{FigPropagatorGammaVertexZ}}
        \quad
        \subfigure[]{\includegraphics[width=0.15\textwidth]{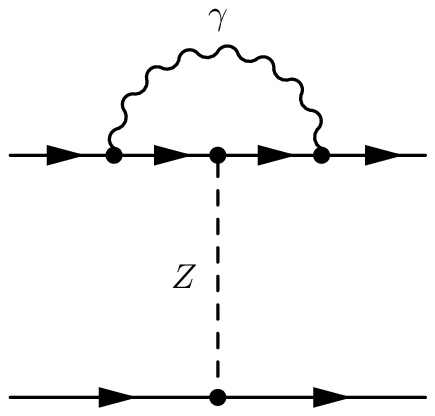}\label{FigPropagatorZVertexG}}
        \quad
        \subfigure[]{\includegraphics[width=0.15\textwidth]{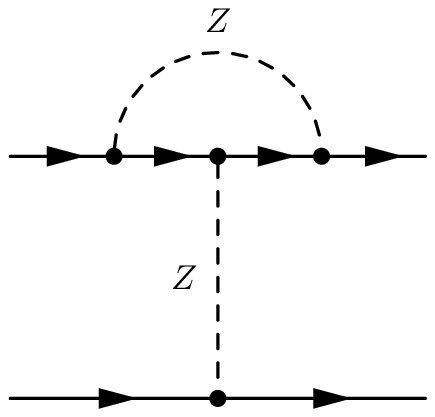}\label{FigPropagatorZVertexZ}}
    }
    \\
    \mbox{
        \subfigure[]{\includegraphics[width=0.15\textwidth]{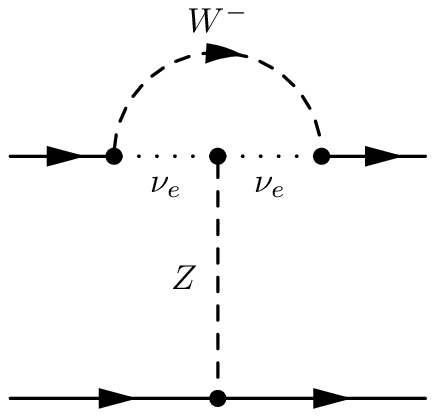}\label{FigPropagatorZVertexW}}
        \quad
        \subfigure[]{\includegraphics[width=0.15\textwidth]{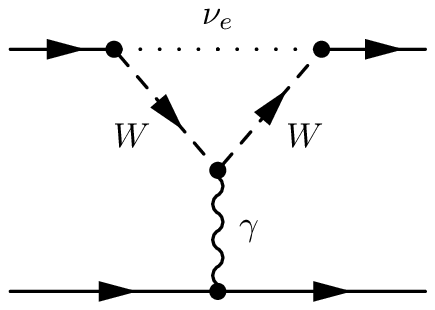}\label{FigPropagatorGVertexW}}
    }
    \caption{One-loop vertex correction diagrams.}
    \label{Fig:OneLoopVertexCorrection}
\end{figure}

It also is necessary to take into account the
contribution of the vertex function of the electron coming from the $W$- and $Z$-boson exchange
(see Fig.~\ref{Fig:OneLoopVertexCorrection}).
The diagrams of Fig.~\ref{FigPropagatorGammaVertexG} and Fig.~\ref{FigPropagatorZVertexG}
are YFS--reducible and should be omitted to avoid double-counting.
It is important to note that in comparison to the contribution from Fig.~\ref{FigPropagatorGammaVertexZ}
and \ref{FigPropagatorGVertexW}, the contributions from Fig.~\ref{FigPropagatorZVertexZ} and \ref{FigPropagatorZVertexW}
contain an additional factor of $\br{t/M_Z^2}$.
Using explicit expressions for the corresponding
contributions to the vertex functions given in Appendix~\ref{AppendixVertexFunction},
we obtain the following contributions to the asymmetry:
\eq{
    A_Z^\Gamma &=
    \frac{\alpha}{\pi}  \bar A_0 \frac{s}{12 M_Z^2}
    \frac{a_V}{c_W^2 s_W^2}
    \br{
        \ln{\frac{M_Z^4}{t u}} + \frac{17}{3}
    },
    \\
    A_W^\Gamma &=
    -     \frac{\alpha}{\pi} \bar A_0 \frac{s}{M_Z^2}
    \frac{1}{c_W^2 s_W^2} \frac{34}{9}.
}
Again, the asymptotic relative correction to the $Z$-boson vertex function
 obtained in \cite{Aleksejevs:2010nf} agrees with the result presented here:
$A_Z^\Gamma \approx \delta^{\Lambda_2}_A A^{(0)}$.

% =========================================================================
\section{Two-loop box radiative corrections}
\label{SectionTwoLoopsRC}
% =========================================================================

It is convenient to divide the two-loop box contributions discussed here into three types:
one type including boxes with lepton self-energy diagrams, and
two other types corresponding to the ladder and decorated-box type.

% ==========================================================================
\subsection{Boxes with lepton self-energy }
\label{SectionHeavySelfEnergy}
% ==========================================================================

For fermions with polarization $\lambda=\pm $, the mass operator has the form
\cite{RQT2}:
\eq{
    \M^e_-(p)\omega_-&=-\frac{p^2\hat{p}}{8\pi^2}\omega_-\brs{\br{\frac{g(1-a_V)}{4c_W}}^2J_Z+\frac{g^2}{2}J_W}, \nn \\
    \M^e_+(p)\omega_+&=-\frac{p^2\hat{p}}{8\pi^2}\omega_+\br{\frac{g(1+a_V)}{4c_W}}^2J_Z, \nn \\
    \M^\nu_-(p)\omega_-&=-\frac{p^2\hat{p}}{8\pi^2}\omega_-\frac{g^2}{2}J_W,
}
where, based on the approach developed in \cite{RQT2} for the pure QED case,
\eq{
    J_Z&=J_Z(p^2)=\int\limits_0^1 d x \int\limits_0^1 dz \frac{x(1-x)}{M_Z^2-p^2x z}, \nn \\
    J_W&=J_Z(M_Z \to M_W).
}

\begin{figure}
    \centering
    \mbox{
        \subfigure[]{\includegraphics[width=0.15\textwidth]{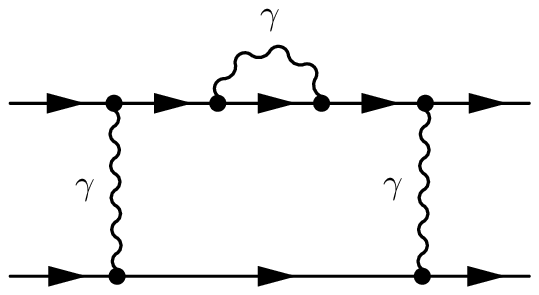}\label{Figggg}}
        \quad
        \subfigure[]{\includegraphics[width=0.15\textwidth]{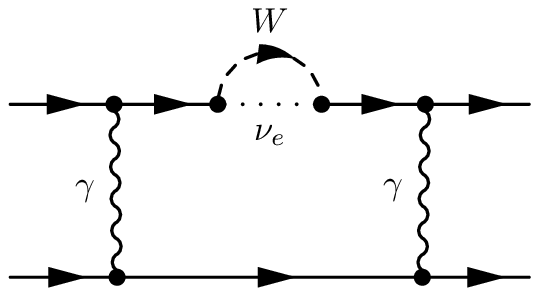}\label{FigggW}}
        \quad
        \subfigure[]{\includegraphics[width=0.15\textwidth]{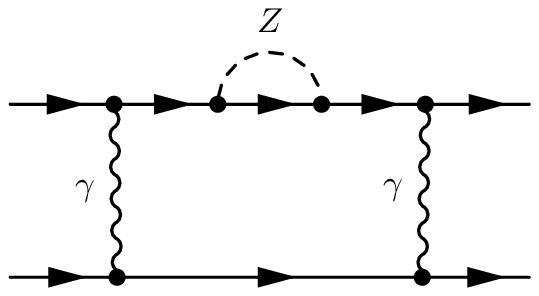}\label{FigggZ}}
        \quad
        \subfigure[]{\includegraphics[width=0.15\textwidth]{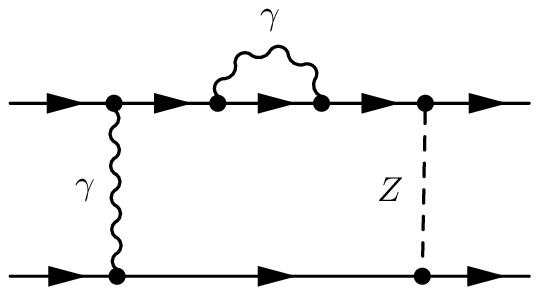}\label{FiggZg}}
        \quad
        \subfigure[]{\includegraphics[width=0.15\textwidth]{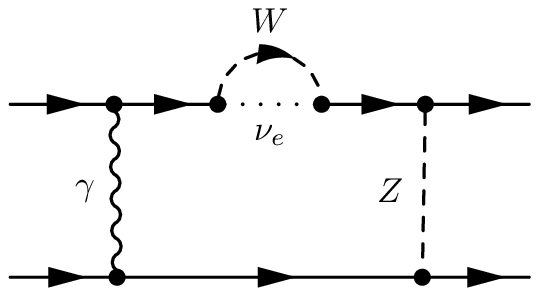}\label{FiggZW}}
    }
    \\
    \mbox{
        \subfigure[]{\includegraphics[width=0.15\textwidth]{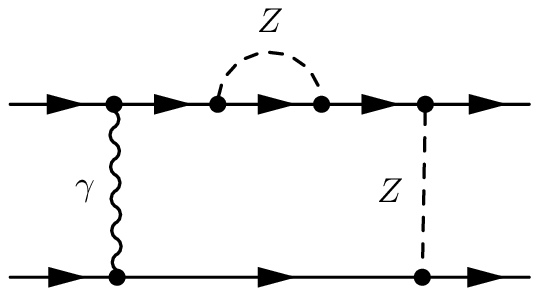}\label{FiggZZ}}
        \quad
        \subfigure[]{\includegraphics[width=0.15\textwidth]{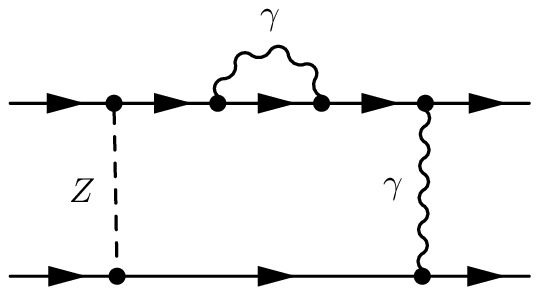}\label{FigZgg}}
        \quad
        \subfigure[]{\includegraphics[width=0.15\textwidth]{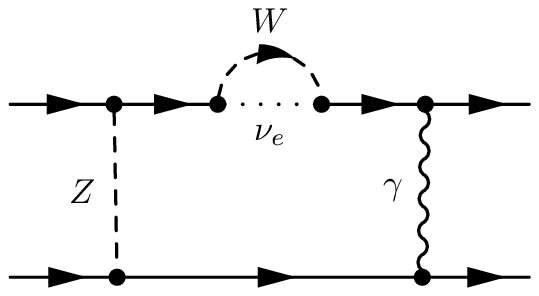}\label{FigZgW}}
        \quad
        \subfigure[]{\includegraphics[width=0.15\textwidth]{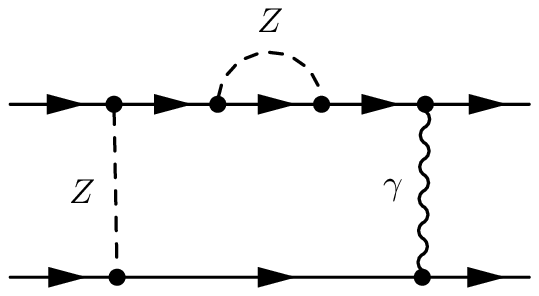}\label{FigZgZ}}
        \quad
        \subfigure[]{\includegraphics[width=0.15\textwidth]{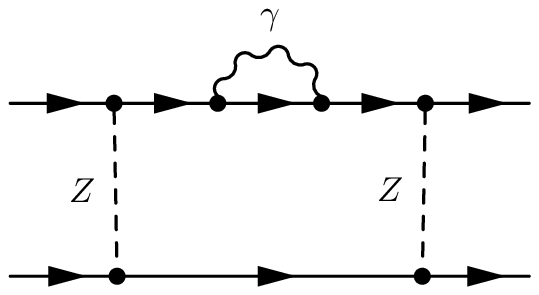}\label{FigZZg}}
        }
    \\
    \mbox{
        \subfigure[]{\includegraphics[width=0.15\textwidth]{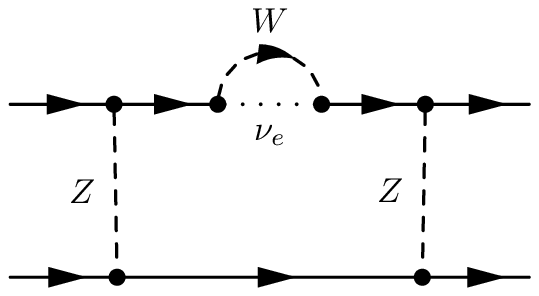}\label{FigZZW}}
        \quad
        \subfigure[]{\includegraphics[width=0.15\textwidth]{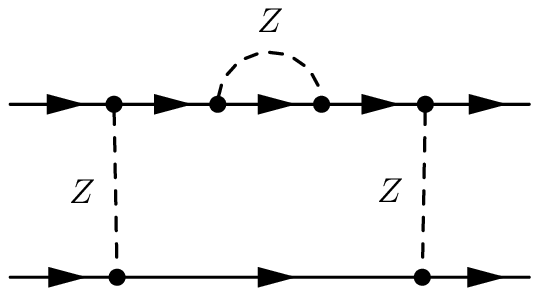}\label{FigZZZ}}
    }
    \caption{One-loop direct box type diagrams with lepton self energy corrections.}
    \label{Fig:OneLoopBoxWithSelfEvergyCorrections}
\end{figure}

\begin{figure}
    \centering
    \mbox{
        \subfigure[]{\includegraphics[width=0.15\textwidth]{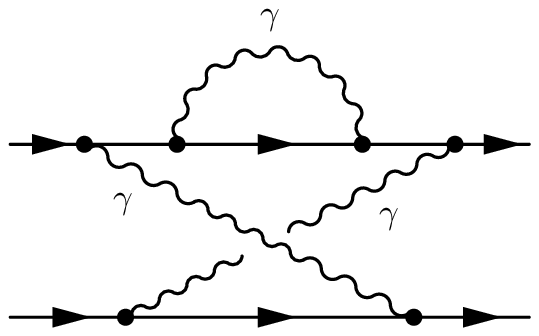}\label{Figggg2}}
        \quad
        \subfigure[]{\includegraphics[width=0.15\textwidth]{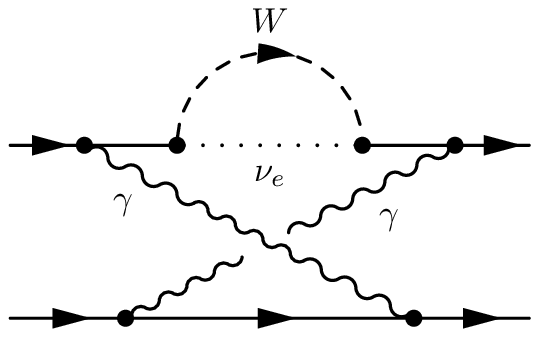}\label{FigggW2}}
        \quad
        \subfigure[]{\includegraphics[width=0.15\textwidth]{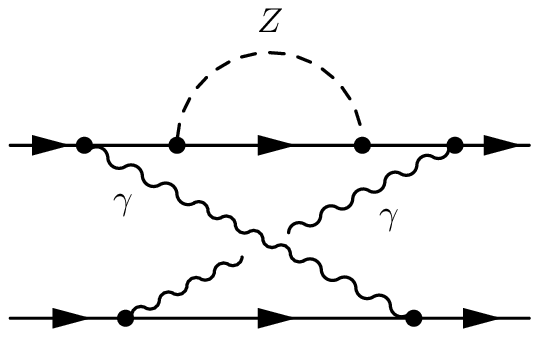}\label{FigggZ2}}
        \quad
        \subfigure[]{\includegraphics[width=0.15\textwidth]{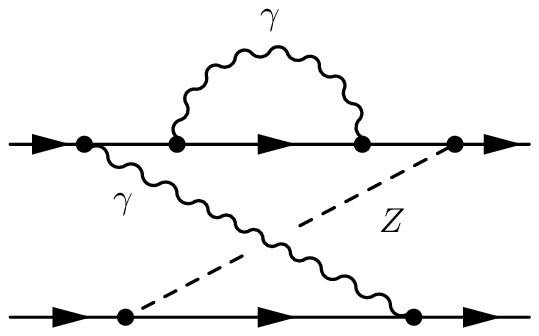}\label{FiggZg2}}
        \quad
        \subfigure[]{\includegraphics[width=0.15\textwidth]{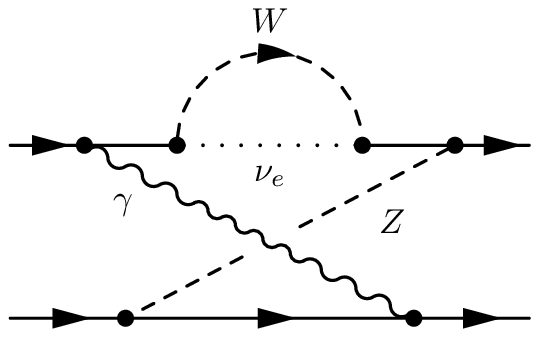}\label{FiggZW2}}
    }
    \\
    \mbox{
        \subfigure[]{\includegraphics[width=0.15\textwidth]{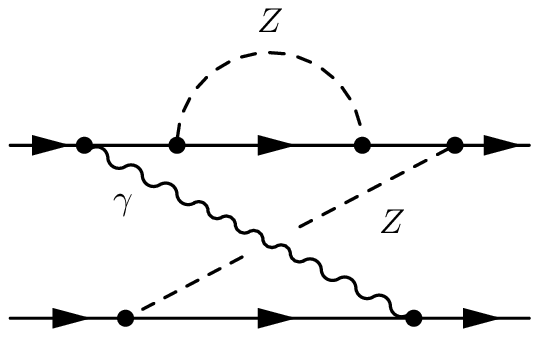}\label{FiggZZ2}}
        \quad
        \subfigure[]{\includegraphics[width=0.15\textwidth]{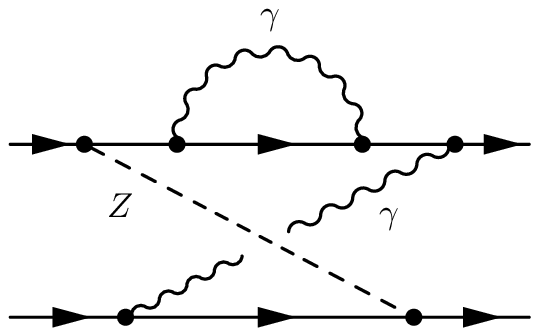}\label{FigZgg2}}
        \quad
        \subfigure[]{\includegraphics[width=0.15\textwidth]{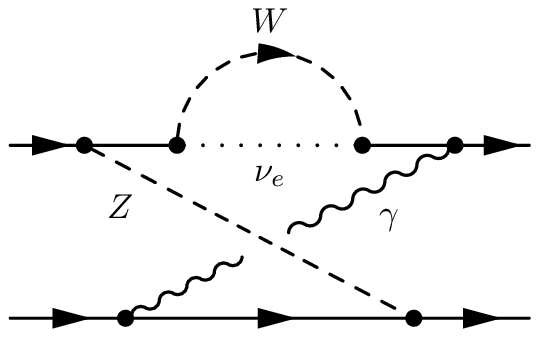}\label{FigZgW2}}
        \quad
        \subfigure[]{\includegraphics[width=0.15\textwidth]{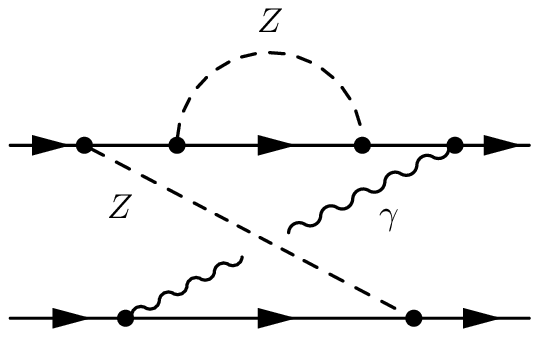}\label{FigZgZ2}}
        \quad
        \subfigure[]{\includegraphics[width=0.15\textwidth]{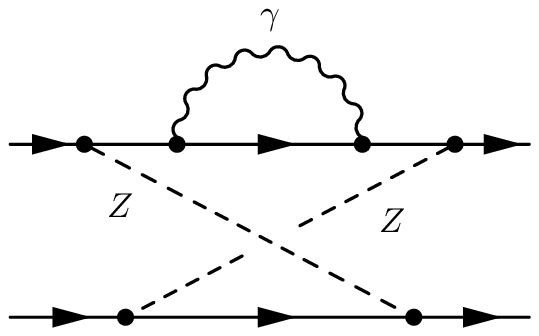}\label{FigZZg2}}
        }
    \\
    \mbox{
        \subfigure[]{\includegraphics[width=0.15\textwidth]{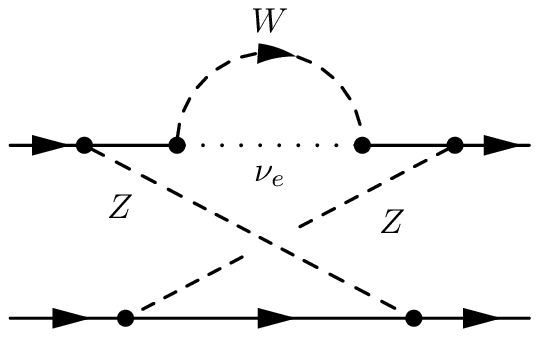}\label{FigZZW2}}
        \quad
        \subfigure[]{\includegraphics[width=0.15\textwidth]{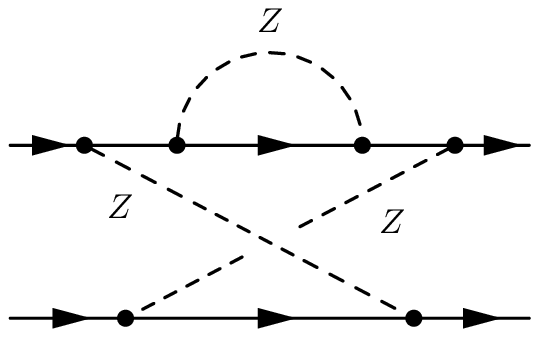}\label{FigZZZ2}}
        \quad
        \subfigure[]{\includegraphics[width=0.15\textwidth]{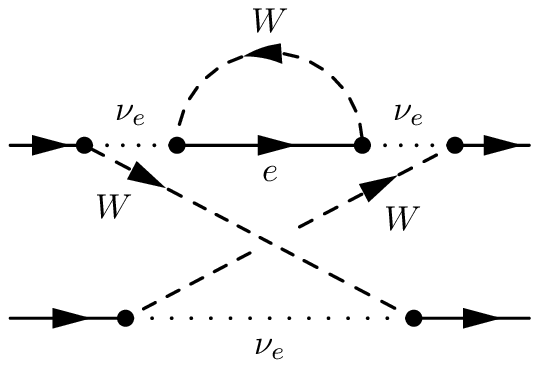}\label{FigWWW2}}
        \quad
        \subfigure[]{\includegraphics[width=0.15\textwidth]{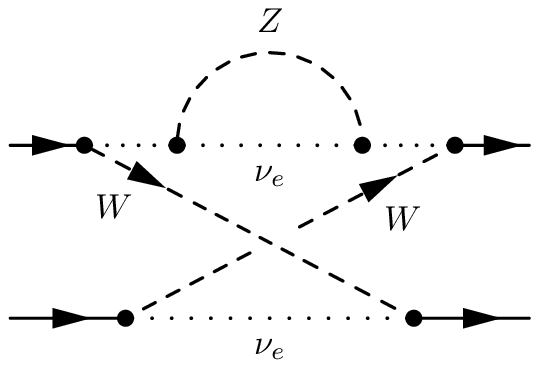}\label{FigWWZ2}}
    }
    \caption{One-loop crossed box type diagrams with lepton self energy corrections.}
    \label{Fig:OneLoopBoxWithSelfEvergyCorrections2}
\end{figure}

Let us evaluate the contribution to the asymmetry coming from self-energy insertions
into the lepton lines ($e$, $\nu$) in the two-boson
$\gamma\gamma$, $\gamma Z$, $ZZ$, $WW$ exchange amplitudes
(all of them are presented in
Figs. \ref{Fig:OneLoopBoxWithSelfEvergyCorrections}
and \ref{Fig:OneLoopBoxWithSelfEvergyCorrections2})

:
\eq{
    A_\Sigma=A^Z_{\gamma Z}+A^W_{\gamma Z}+A^\nu_{WW}.
}
For $A^Z_{\gamma Z}$, we have
\eq{
    A^Z_{\gamma Z} &= - \br{\frac{\alpha}{\pi}}^2 \bar A_0  \frac{24 s a_V}{M_Z^2} \rho^2  {\cal J}_Z,
    \qquad
    L_Z=\ln\frac{M_Z^2}{m^2},
    \qquad
    \rho = \frac{1}{4 c_W s_W},
    \\
    {\cal J}_Z &=
    \frac{1}{6} L_Z +
    \int\limits_0^1 dx x \bar x
    \int\limits_0^1 dz
    \brf{
        -\ln\beta +
        \rho^2\int\limits_0^\infty
        \frac{dt}{1+\beta t}
        \brs{
            \frac{4}{t+1} + \frac{3\rho^2 t}{\br{t+1}^2}
        }
    }
    =
    \frac{1}{6} L_Z + 0.888,
}
where $\bar x = 1-x$ and $\beta = x z$.
The intermediate $W$ state in the electron self-energy gives
\eq{
    A^W_{\gamma Z} &= \br{\frac{\alpha}{\pi}}^2  \bar A_0 \frac{3s}{M_Z^2} \frac{1}{s_W^2} {\cal J}_W,
    \qquad
    L_W=\ln\frac{M_W^2}{m^2},
    \qquad
    a = \frac{1}{c_W^2},
    \\
    {\cal J}_W
    &=
    \frac{1}{6} L_W +
    \int\limits_0^1 dx x \bar x
    \int\limits_0^1 dz
    \brf{
        -\ln\beta +
        \rho^2\int\limits_0^\infty
        \frac{dt}{1+\beta t}
        \brs{
            \frac{2}{t+a} + \frac{\rho^2}{\br{t+a}^2}
        }
    }
    =
    \frac{1}{6} L_W + 0.542.
}
The contribution from the neutrino self-energy insertion is
\eq{
    A^\nu_{WW}= \br{\frac{\alpha}{\pi}}^2 \bar A_0 \frac{s}{8 M_Z^2}
    \frac{1}{s_W^6 c_W^2} R^{\nu},
}
where
\eq{
    R^\nu &=\int\limits_0^1 d x \, x \bar x\int\limits_0^1 d z \int\limits_0^\infty
    \frac{t d t}{(t+1)^2}\br{\frac{1}{a+\beta t}+\frac{2 c_W^2}{1+\beta t}} = 0.586.
}

%(Corresponding diagrams would be great here.) ==========================================================================
\subsection{Ladder-box diagrams}
\label{SectionDoubleBox}
% ==========================================================================

% ==========================================================================
\subsubsection{$ZZZ$ exchange}
\label{SectionDoubleBoxZZZ}
% ==========================================================================
\begin{figure}
    \centering
    \mbox{
        \subfigure[]{\includegraphics[width=0.15\textwidth]{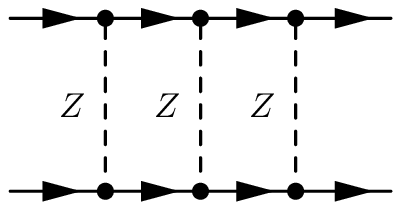}\label{FigDoubleBoxZ1Z2Z3}}
        \quad
        \subfigure[]{\includegraphics[width=0.15\textwidth]{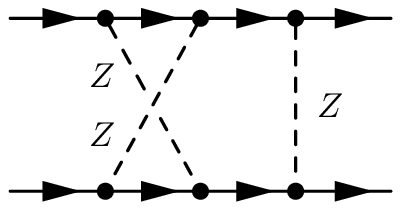}\label{FigDoubleBoxZ1Z3Z2}}
        \quad
        \subfigure[]{\includegraphics[width=0.15\textwidth]{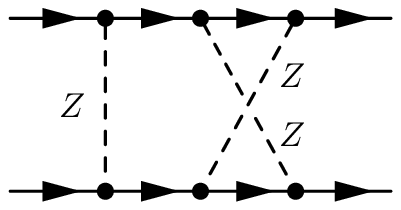}\label{FigDoubleBoxZ2Z1Z3}}
    }
    \\
    \mbox{
        \subfigure[]{\includegraphics[width=0.15\textwidth]{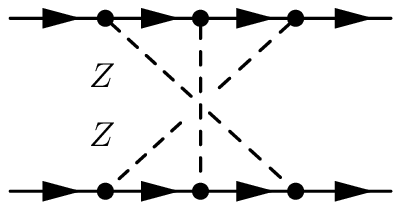}\label{FigDoubleBoxZ3Z2Z1}}
        \quad
        \subfigure[]{\includegraphics[width=0.15\textwidth]{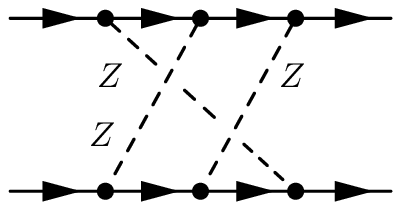}\label{FigDoubleBoxZ3Z1Z2}}
        \quad
        \subfigure[]{\includegraphics[width=0.15\textwidth]{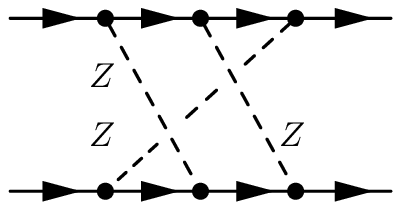}\label{FigDoubleBoxZ2Z3Z1}}
    }
    \caption{Double-box-type diagrams with $ZZZ$ exchange (see Section~\ref{SectionDoubleBoxZZZ}).}
    \label{Fig:DoubleBoxZZZ}
\end{figure}

The contribution of the ladder-box diagrams with an exchange of three $Z$ bosons
(Figs.~\ref{FigDoubleBoxZ1Z2Z3}-\ref{FigDoubleBoxZ2Z3Z1})
to the asymmetry has the form:
\eq{
    A_{ZZZ}
    &=
    \br{\frac{\alpha}{\pi}}^2 \bar A_0
    \frac{3 s a_V}{M_Z^2} \rho^6
    N_{ZZZ},
}
where $N_{ZZZ}$ is the loop momentum integral:
\eq{
    N_{ZZZ} = N_{123}^{ZZZ} + N_{132}^{ZZZ} + N_{213}^{ZZZ} + N_{321}^{ZZZ} + N_{312}^{ZZZ} + N_{213}^{ZZZ}
    = 2.40755,
}
which includes 6 terms corresponding to 6
diagrams with three $Z$-boson exchanges:
\eq{
    &\text{Fig.~\ref{FigDoubleBoxZ1Z2Z3}}:
    &N_{123}^{ZZZ} &= \int \frac{d\chi \, M_Z^2}{\a_e^2\,\b_e^2\,\a_Z\,\b_Z\,\c_Z} S_{123} = -59.8697,
    \\
    &\text{Fig.~\ref{FigDoubleBoxZ1Z3Z2}}:
    &N_{132}^{ZZZ} &= \int \frac{d\chi \, M_Z^2}{\a_e\,\b_e^2\,\c_e\,\a_Z\,\b_Z\,\c_Z} S_{132} = 16.6497,
    \\
    &\text{Fig.~\ref{FigDoubleBoxZ2Z1Z3}}:
    &N_{213}^{ZZZ} &= \int \frac{d\chi \, M_Z^2}{\a_e^2\,\b_e\,\c_e\,\a_Z\,\b_Z\,\c_Z} S_{213} = 16.6497,
    \\
    &\text{Fig.~\ref{FigDoubleBoxZ3Z2Z1}}:
    &N_{321}^{ZZZ} &= \int \frac{d\chi \, M_Z^2}{\a_e^2\,\b_e^2\,\a_Z\,\b_Z\,\c_Z} S_{321} = -4.32159,
    \\
    &\text{Fig.~\ref{FigDoubleBoxZ3Z1Z2}}:
    &N_{312}^{ZZZ} &= \int \frac{d\chi \, M_Z^2}{\a_e^2\,\b_e\,\c_e\,\a_Z\,\b_Z\,\c_Z} S_{312} = 16.6497,
    \\
    &\text{Fig.~\ref{FigDoubleBoxZ2Z3Z1}}:
    &N_{231}^{ZZZ} &= \int \frac{d\chi \, M_Z^2}{\a_e\,\b_e^2\,\c_e\,\a_Z\,\b_Z\,\c_Z} S_{231} = 16.6497.
}
Here we use traces and notations defined in Appendix~\ref{AppendixTwoLoopsTopologiesAndTraces}.
The loop momentum integrals and notations for the propagator denominators
are defined in Appendix~\ref{AppendixTwoLoopMomentumIntegration}.

% ==========================================================================
\subsubsection{$ZZ\gamma$ and $WW\gamma$ exchange}
\label{SectionDoubleBoxZZGammaWWGamma}
% ==========================================================================
\begin{figure}
    \centering
    \mbox{
        \subfigure[]{\includegraphics[width=0.15\textwidth]{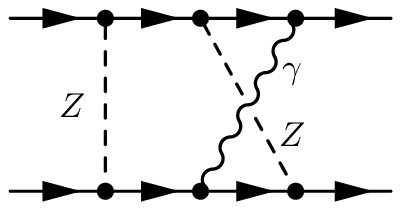}\label{FigDoubleBoxZ2g1Z3}}
        \quad
        \subfigure[]{\includegraphics[width=0.15\textwidth]{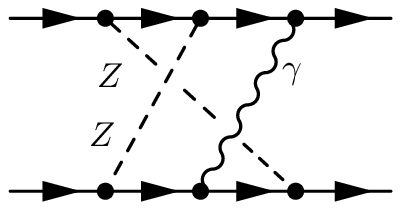}\label{FigDoubleBoxZ3g1Z2}}
        \quad
        \subfigure[]{\includegraphics[width=0.15\textwidth]{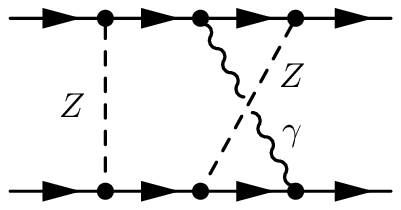}\label{FigDoubleBoxg2Z1Z3}}
        \quad
        \subfigure[]{\includegraphics[width=0.15\textwidth]{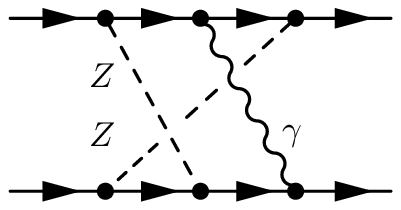}\label{FigDoubleBoxg2Z3Z1}}
        \quad
        \subfigure[]{\includegraphics[width=0.15\textwidth]{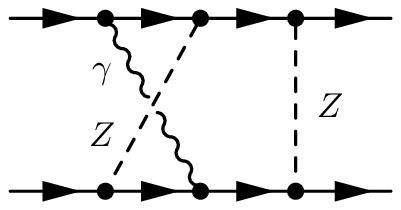}\label{FigDoubleBoxZ1g3Z2}}
    }
    \\
    \mbox{
        \subfigure[]{\includegraphics[width=0.15\textwidth]{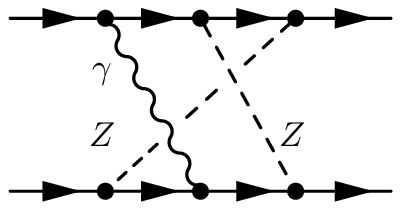}\label{FigDoubleBoxZ2g3Z1}}
        \quad
        \subfigure[]{\includegraphics[width=0.15\textwidth]{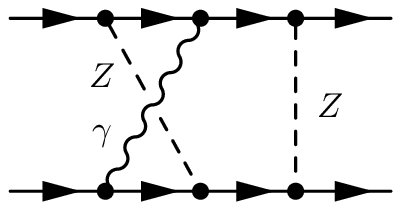}\label{FigDoubleBoxZ1Z3g2}}
        \quad
        \subfigure[]{\includegraphics[width=0.15\textwidth]{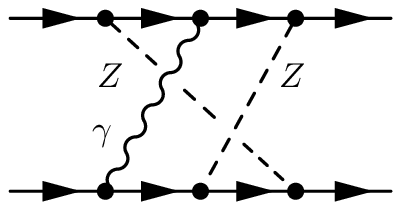}\label{FigDoubleBoxZ3Z1g2}}
        \quad
        \subfigure[]{\includegraphics[width=0.15\textwidth]{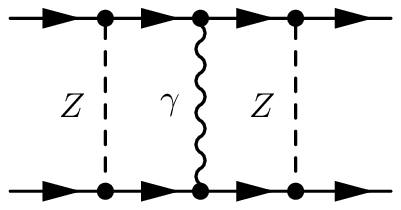}\label{FigDoubleBoxZ1g2Z3}}
        \quad
        \subfigure[]{\includegraphics[width=0.15\textwidth]{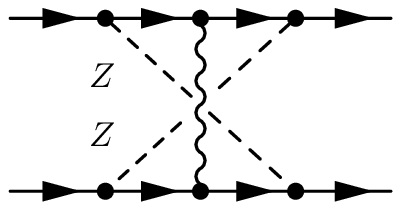}\label{FigDoubleBoxZ3g2Z1}}
    }
    \caption{Double-box-type diagrams with $ZZ\gamma$ exchange (see Section~\ref{SectionDoubleBoxZZGammaWWGamma}).}
    \label{Fig:DoubleBoxZZGamma}
\end{figure}

The contribution of the double-box diagrams with one photon and two $Z$-boson exchanges
(Fig.~\ref{Fig:DoubleBoxZZGamma})
to the asymmetry has the form:
\eq{
    A_{ZZ\gamma}
    &=
    - \br{\frac{\alpha}{\pi}}^2 \bar A_0
    \frac{2 s a_V}{M_Z^2} \rho^4
    N_{ZZ\gamma},
}
where $N_{ZZ\gamma}$ is the loop momentum integral
\eq{
    N_{ZZ\gamma} &=
    N_{\Z{2}\g{1}\Z{3}}^{ZZ\gamma} + N_{\Z{3}\g{1}\Z{2}}^{ZZ\gamma} + N_{\g{2}\Z{1}\Z{3}}^{ZZ\gamma} +
    N_{\g{2}\Z{3}\Z{1}}^{ZZ\gamma} + N_{\Z{1}\g{3}\Z{2}}^{ZZ\gamma} + N_{\Z{2}\g{3}\Z{1}}^{ZZ\gamma} +
    N_{\Z{1}\Z{3}\g{2}}^{ZZ\gamma} + N_{\Z{3}\Z{1}\g{2}}^{ZZ\gamma} + N_{\Z{1}\g{2}\Z{3}}^{ZZ\gamma} +
    N_{\Z{3}\g{2}\Z{1}}^{ZZ\gamma}
    =\nn\\
    &= 8 L_Z + 61.2176,
}
which includes 10 terms corresponding to 10
diagrams with one photon and two $Z$-boson exchanges:
\eq{
    &\text{Fig.~\ref{FigDoubleBoxZ2g1Z3}}:
    &N_{\Z{2}\g{1}\Z{3}}^{ZZ\gamma} &= \int \frac{d\chi \, M_Z^2}{\a_e^2\,\b_e\,\c_e\,\a_Z\,\b\,\c_Z} S_{213} = 13.1595,
    \\
    &\text{Fig.~\ref{FigDoubleBoxZ3g1Z2}}:
    &N_{\Z{3}\g{1}\Z{2}}^{ZZ\gamma} &= \int \frac{d\chi \, M_Z^2}{\a_e^2\,\b_e\,\c_e\,\a_Z\,\b\,\c_Z} S_{312} = 13.1595,
    \\
    &\text{Fig.~\ref{FigDoubleBoxg2Z1Z3}}:
    &N_{\g{2}\Z{1}\Z{3}}^{ZZ\gamma} &= \int \frac{d\chi \, M_Z^2}{\a_e^2\,\b_e\,\c_e\,\a_Z\,\b_Z\,\c} S_{213} = 39.4784,
    \\
    &\text{Fig.~\ref{FigDoubleBoxg2Z3Z1}}:
    &N_{\g{2}\Z{3}\Z{1}}^{ZZ\gamma} &= \int \frac{d\chi \, M_Z^2}{\a_e\,\b_e^2\,\c_e\,\a_Z\,\b_Z\,\c} S_{231} = 4 L_Z + 19.7392,
    \\
    &\text{Fig.~\ref{FigDoubleBoxZ1g3Z2}}:
    &N_{\Z{1}\g{3}\Z{2}}^{ZZ\gamma} &= \int \frac{d\chi \, M_Z^2}{\a_e\,\b_e^2\,\c_e\,\a\,\b_Z\,\c_Z} S_{132} = 13.1595,
    \\
    &\text{Fig.~\ref{FigDoubleBoxZ2g3Z1}}:
    &N_{\Z{2}\g{3}\Z{1}}^{ZZ\gamma} &= \int \frac{d\chi \, M_Z^2}{\a_e\,\b_e^2\,\c_e\,\a\,\b_Z\,\c_Z} S_{231} = 13.1595,
    \\
    &\text{Fig.~\ref{FigDoubleBoxZ1Z3g2}}:
    &N_{\Z{1}\Z{3}\g{2}}^{ZZ\gamma} &= \int \frac{d\chi \, M_Z^2}{\a_e\,\b_e^2\,\c_e\,\a_Z\,\b_Z\,\c} S_{132} = 4 L_Z + 19.7392,
    \\
    &\text{Fig.~\ref{FigDoubleBoxZ3Z1g2}}:
    &N_{\Z{3}\Z{1}\g{2}}^{ZZ\gamma} &= \int \frac{d\chi \, M_Z^2}{\a_e^2\,\b_e\,\c_e\,\a_Z\,\b_Z\,\c} S_{312} = 39.4784,
    \\
    &\text{Fig.~\ref{FigDoubleBoxZ1g2Z3}}:
    &N_{\Z{1}\g{2}\Z{3}}^{ZZ\gamma} &= \int \frac{d\chi \, M_Z^2}{\a_e^2\,\b_e^2\,\a_Z\,\b_Z\,\c} S_{123} = -105.276,
    \\
    &\text{Fig.~\ref{FigDoubleBoxZ3g2Z1}}:
    &N_{\Z{3}\g{2}\Z{1}}^{ZZ\gamma} &= \int \frac{d\chi \, M_Z^2}{\a_e^2\,\b_e^2\,\a_Z\,\b_Z\,\c} S_{321} = -4.57974.
}
As before, we use traces and notations defined in Appendix~\ref{AppendixTwoLoopsTopologiesAndTraces} and the loop momentum integrals and notations for the propagator denominators
defined in Appendix~\ref{AppendixTwoLoopMomentumIntegration}.
\begin{figure}
        \subfigure[]{\includegraphics[width=0.15\textwidth]{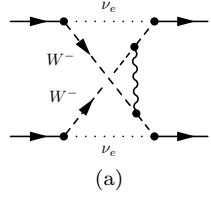}\label{FigLadderWW}}
    \caption{Double-box-type diagrams of ladder form $WW\gamma$ exchange (see Section~\ref{SectionDoubleBoxZZGammaWWGamma}).}
    \label{Fig:LadderBoxWWGamma}
\end{figure}

The contribution of the double-box diagrams of ladder form with one-photon
exchange between two $W$-bosons (Fig.~\ref{Fig:LadderBoxWWGamma}) is given by:
\eq{
    A_{WW\gamma}
    &=
    - \br{\frac{\alpha}{\pi}}^2  \bar A_0
    \frac{s}{16 M_Z^2} \frac{1}{s_W^4 c_W^2}
    N_{WW\gamma},
}
where $N_{WW\gamma}$ is the loop momentum integral
\eq{
    N_{WW\gamma}
    &=
    \int \frac{d\chi \, M_W^2}{\a_e\,\c_e\,\a_W^2\,\c_W^2\,\b} S_{WW\gamma} = 11.1595,
}
and the trace $S_{WW\gamma}$ is presented in (\ref{eq:SpWWg}).

% ==========================================================================
\subsubsection{$ZWW$ exchange}
\label{SectionDoubleBoxZWW}
% ==========================================================================
\begin{figure}
    \centering
    \mbox{
        \subfigure[]{\includegraphics[width=0.15\textwidth]{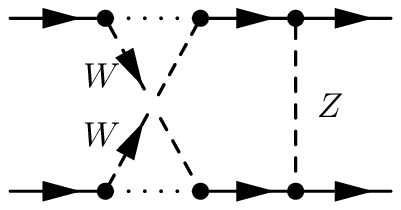}\label{FigDoubleBoxZWW132}}
        \quad
        \subfigure[]{\includegraphics[width=0.15\textwidth]{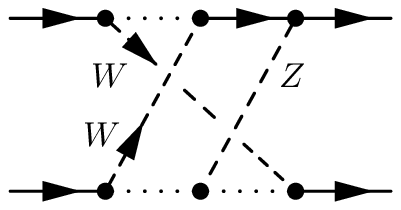}\label{FigDoubleBoxZWWi312}}
        \quad
        \subfigure[]{\includegraphics[width=0.15\textwidth]{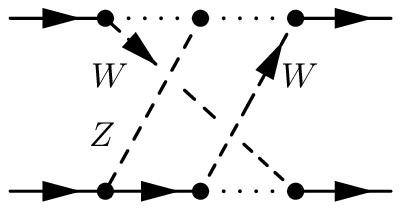}\label{FigDoubleBoxZWWii312}}
        \quad
        \subfigure[]{\includegraphics[width=0.15\textwidth]{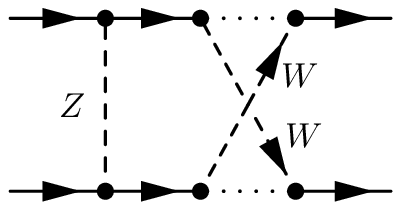}\label{FigDoubleBoxZWW213}}
    }
    \\
    \mbox{
        \subfigure[]{\includegraphics[width=0.15\textwidth]{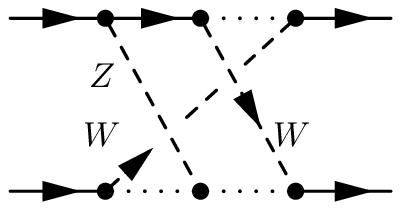}\label{FigDoubleBoxZWWi231}}
        \quad
        \subfigure[]{\includegraphics[width=0.15\textwidth]{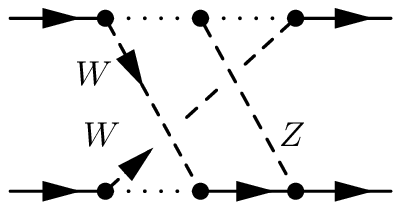}\label{FigDoubleBoxZWWii231}}
        \quad
        \subfigure[]{\includegraphics[width=0.15\textwidth]{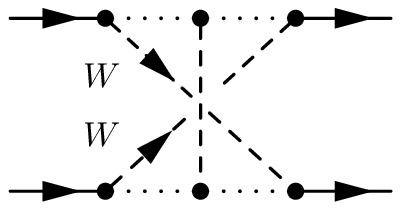}\label{FigDoubleBoxZWW321}}
    }
    \caption{Double-box-type diagrams with $ZWW$ exchange (see Section~\ref{SectionDoubleBoxZWW}).}
    \label{Fig:DoubleBoxZWW}
\end{figure}

The contribution of the double-box diagrams with one photon and two $Z$-bosons exchanges,
presented in Fig.~\ref{Fig:DoubleBoxZWW},
can be expressed as:
\eq{
    A_{ZWW}
    &=
    -  \br{\frac{\alpha}{\pi}}^2 \bar A_0
    \frac{s}{16 M_Z^2} \frac{\rho^2}{s_W^4}
    N_{ZWW},
}
where $N_{ZWW}$ is the loop momentum integral,
\eq{
    N_{ZWW} =
    N_{132}^{ZWW} + N_{312}^{ZWW(1)} + N_{312}^{ZWW(2)} + N_{213}^{ZWW} +
    N_{231}^{ZWW(1)} + N_{231}^{ZWW(2)} + N_{321}^{ZWW}
    = -24.4674,
}
which includes 7 terms corresponding to 7
diagrams with one $Z$-boson and two $W$-boson exchanges:
\eq{
    &\text{Fig.~\ref{FigDoubleBoxZWW132}}:
    &N_{132}^{ZWW} &= \int \frac{d\chi \, M_Z^2}{\b_e^2\,\a_W\,\b_Z\,\c_W\,\a\,\c} S_{132} = 8.52405,
    \\
    &\text{Fig.~\ref{FigDoubleBoxZWWi312}}:
    &N_{312}^{ZWW(1)} &= \int \frac{d\chi \, M_Z^2}{\b_e\,\a_W\,\b_Z\,\c_W\,\a^2\,\c} S_{312} = -27.2451,
    \\
    &\text{Fig.~\ref{FigDoubleBoxZWWii312}}:
    &N_{312}^{ZWW(2)} &= \int \frac{d\chi \, M_Z^2}{\c_e\,\a_W\,\b_W\,\c_Z\,\a^2\,\b} S_{312} = 9.08169,
    \\
    &\text{Fig.~\ref{FigDoubleBoxZWW213}}:
    &N_{213}^{ZWW} &= \int \frac{d\chi \, M_Z^2}{\a_e^2\,\a_Z\,\b_W\,\c_W\,\b\,\c} S_{213} = 8.52405,
    \\
    &\text{Fig.~\ref{FigDoubleBoxZWWi231}}:
    &N_{231}^{ZWW(1)} &= \int \frac{d\chi \, M_Z^2}{\a_e\,\a_Z\,\b_W\,\c_W\,\b^2\,\c} S_{231} = -27.2451,
    \\
    &\text{Fig.~\ref{FigDoubleBoxZWWii231}}:
    &N_{231}^{ZWW(2)} &= \int \frac{d\chi \, M_Z^2}{\c_e\,\a_W\,\b_W\,\c_Z\,\a\,\b^2} S_{231} = 9.08169,
    \\
    &\text{Fig.~\ref{FigDoubleBoxZWW321}}:
    &N_{321}^{ZWW} &= \int \frac{d\chi \, M_Z^2}{\a_W\,\b_W\,\c_Z\,\a^2\,\b^2} S_{321} = -5.18876,
}
where the traces $S_{132,\cdots}$ coincide with the traces from the $ZZZ$ case (see Appendix~\ref{AppendixTwoLoopsTopologiesAndTraces}).

% ==========================================================================
\subsection{Decorated-box diagrams}
\label{SectionDecoratedBox}
% ==========================================================================

% ==========================================================================
\subsubsection{Type I}
\label{SectionDecoratedBoxTypeI}
% ==========================================================================
\begin{figure}
    \centering
    \mbox{
        \subfigure[]{\includegraphics[width=0.15\textwidth]{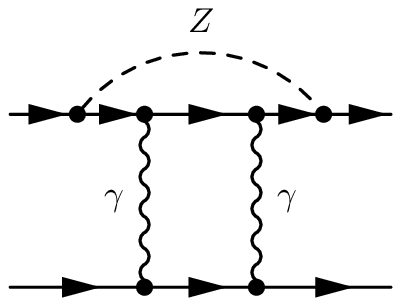}\label{FigDecoratedBoxGGOveralZ1}}
        \quad
        \subfigure[]{\includegraphics[width=0.15\textwidth]{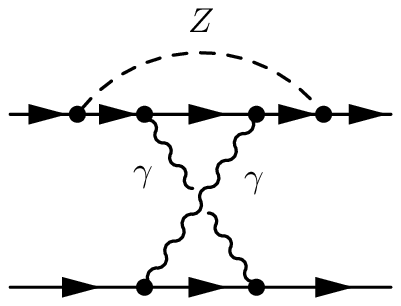}\label{FigDecoratedBoxGGOveralZ2}}
        \quad
        \subfigure[]{\includegraphics[width=0.15\textwidth]{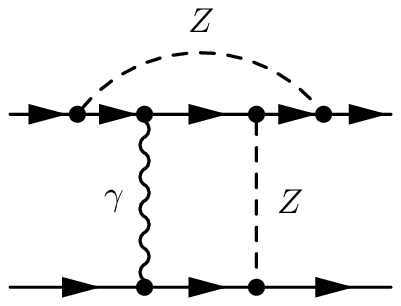}\label{FigDecoratedBoxGZOveralZ1}}
        \quad
        \subfigure[]{\includegraphics[width=0.15\textwidth]{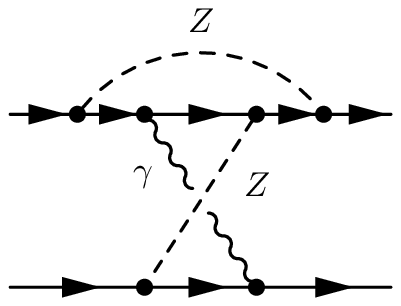}\label{FigDecoratedBoxGZOveralZ2}}
    }
    \\
    \mbox{
        \subfigure[]{\includegraphics[width=0.15\textwidth]{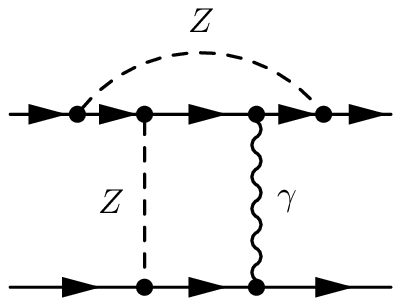}\label{FigDecoratedBoxZGOveralZ1}}
        \quad
        \subfigure[]{\includegraphics[width=0.15\textwidth]{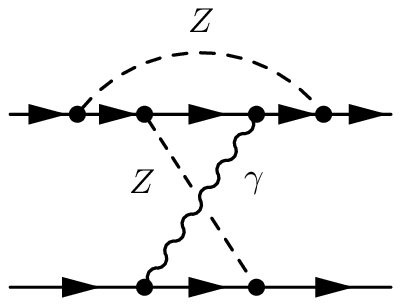}\label{FigDecoratedBoxZGOveralZ2}}
        \quad
        \subfigure[]{\includegraphics[width=0.15\textwidth]{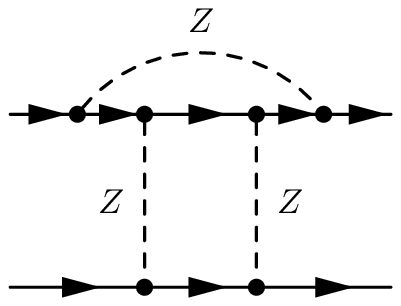}\label{FigDecoratedBoxZZOveralZ1}}
        \quad
        \subfigure[]{\includegraphics[width=0.15\textwidth]{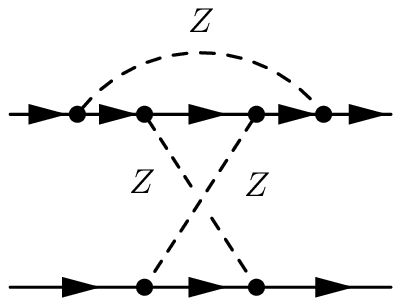}\label{FigDecoratedBoxZZOveralZ2}}
    }
    \\
    \mbox{
        \subfigure[]{\includegraphics[width=0.15\textwidth]{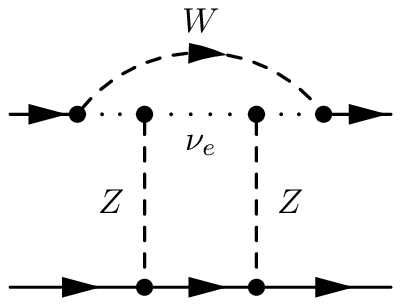}\label{FigDecoratedBoxZZOveralW1}}
        \quad
        \subfigure[]{\includegraphics[width=0.15\textwidth]{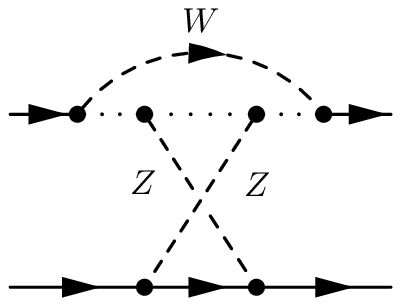}\label{FigDecoratedBoxZZOveralW2}}
        \quad
        \subfigure[]{\includegraphics[width=0.15\textwidth]{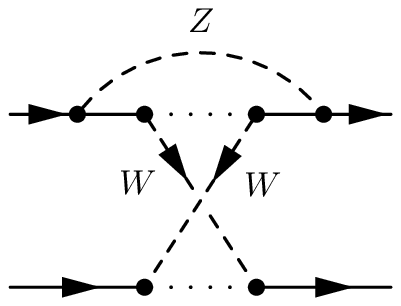}\label{FigDecoratedBoxWWOveralZ}}
        }
    \caption{Decorated box diagrams of Type I (see Section~\ref{SectionDecoratedBoxTypeI}).}
    \label{Fig:DecoratedBoxDiagramsTypeI}
\end{figure}

The contribution to the asymmetry coming from the decorated-box diagrams with two photon exchanges,
presented in Fig.~\ref{FigDecoratedBoxGGOveralZ1} and \ref{FigDecoratedBoxGGOveralZ2},
has the form:
\eq{
    A_{\gamma\gamma}^Z
    &=
    \br{\frac{\alpha}{\pi}}^2 \bar A_0
    \frac{2 s a_V}{M_Z^2} \rho^2
    N_{\gamma\gamma}^Z,
}
where $N_{\gamma\gamma}^Z$ is the loop momentum integral:
\eq{
    N_{\gamma\gamma}^Z
    &=
    \int \frac{d\chi \, M_Z^2}{\a_e^2\,\b_e^2\,\c_e\,\a_Z\,\b^2}
    \br{ S_1^I + S_2^I }
    =
    \brs{
        2 L_Z^2 + 6 L_Z + 13.1595
    }
    +
    \brs{
        -2 L_Z^2 -13.6595
    }
    =6 L_Z - 0.5,
}
and the traces $S_{1,2}^I$ are defined in Appendix~\ref{AppendixTwoLoopsTopologiesAndTraces}.
The contribution of decorated-box diagrams with one photon and one $Z$-boson exchange
(Fig.~\ref{FigDecoratedBoxGZOveralZ1}, \ref{FigDecoratedBoxGZOveralZ2}, \ref{FigDecoratedBoxZGOveralZ1}
and \ref{FigDecoratedBoxZGOveralZ2}) is given by:
\eq{
    A_{\gamma Z}^Z
    =
    - \br{\frac{\alpha}{\pi}}^2 \bar A_0
    \frac{16 s a_V}{M_Z^2} \rho^4
    N_{\gamma Z}^Z,
    \qquad
    N_{\gamma Z}^Z
    =
    \int \frac{d\chi \, M_Z^2}{\a_e^2\,\b_e^2\,\c_e\,\a_Z\,\b_Z\,\b}
    \br{ S_1^I + S_2^I }
%    =
%    17.1595 + (-8.57974) + 17.1595 + (-8.57974)
%    = 17.1595/2
    = 8.57974.
}
The decorated-box diagrams with two $Z$-boson exchanges
(Fig.~\ref{FigDecoratedBoxZZOveralZ1} and \ref{FigDecoratedBoxZZOveralZ2}) contribute the following:
\eq{
    A_{ZZ}^Z
    =
    - \br{\frac{\alpha}{\pi}}^2 \bar A_0
    \frac{6 s a_V}{M_Z^2} \rho^6
    N_{ZZ}^Z,
    \qquad
    N_{ZZ}^Z
    =
    \int \frac{d\chi \, M_Z^2}{\a_e^2\,\b_e\,\c_e\,\a_Z\,\b_Z^2} \br{ S_1^I + S_2^I }
%    =
%    9.15947 + (-4.)
    = 5.15947.
}
The decorated-box diagrams with two $Z$-boson exchanges
(Fig.~\ref{FigDecoratedBoxZZOveralW1} and \ref{FigDecoratedBoxZZOveralW2} give the term:
\eq{
    A_{ZZ}^W
    =
    - \br{\frac{\alpha}{\pi}}^2 \bar A_0
    \frac{s}{M_Z^2} \frac{\rho^4}{s_W^2}
    N_{ZZ}^W,
    \qquad
    N_{ZZ}^W
    =
    \int \frac{d\chi \, M_Z^2}{\a^2\,\b_e\,\c\,\a_W\,\b_Z^2} \br{ S_1^I + S_2^I }
%    = 10.251 + (-4.3822)
    = 5.86885.
}
And, finally, the contribution to the asymmetry from the decorated-box diagrams with two $W$-boson exchanges,
presented in Fig.~\ref{FigDecoratedBoxWWOveralZ}, has the form:
\eq{
    A_{WW}^Z
    =
    - \br{\frac{\alpha}{\pi}}^2 \bar A_0
    \frac{s}{M_Z^2} \frac{\rho^2}{8 s_W^4}
    N_{WW}^Z,
    \qquad
    N_{WW}^Z
    =
    \int \frac{d\chi \, M_Z^2}{\a_e^2\,\b\,\c\,\a_Z\,\b_W^2} S_2^I
    = -11.7914.
}

% ==========================================================================
\subsubsection{Type II}
\label{SectionDecoratedBoxTypeII}
% ==========================================================================
%
\begin{figure}
    \centering
    \mbox{
        \subfigure[]{\includegraphics[width=0.15\textwidth]{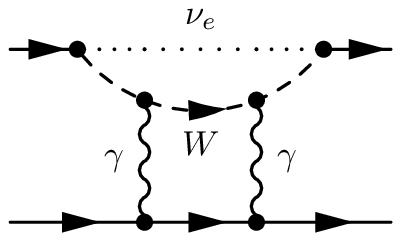}\label{FigDecoratedBoxGGUnderalW1}}
        \quad
        \subfigure[]{\includegraphics[width=0.15\textwidth]{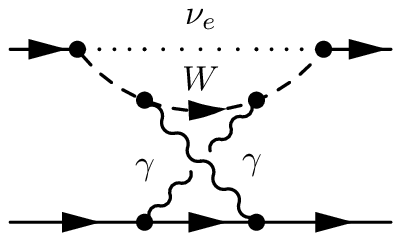}\label{FigDecoratedBoxGGUnderalW2}}
        \quad
        \subfigure[]{\includegraphics[width=0.15\textwidth]{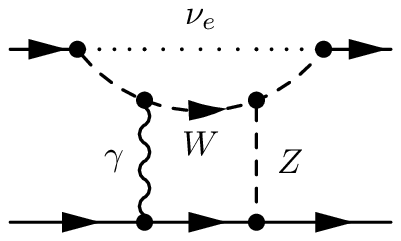}\label{FigDecoratedBoxGZUnderalW1}}
        \quad
        \subfigure[]{\includegraphics[width=0.15\textwidth]{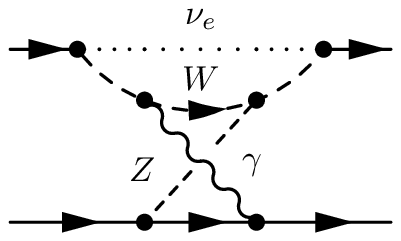}\label{FigDecoratedBoxGZUnderalW2}}
        \quad
        \subfigure[]{\includegraphics[width=0.15\textwidth]{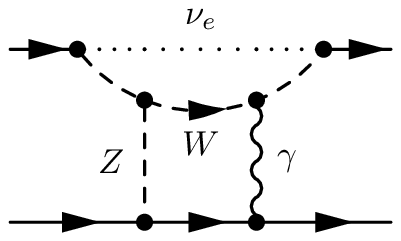}\label{FigDecoratedBoxZGUnderalW1}}
    }
    \\
    \mbox{
        \subfigure[]{\includegraphics[width=0.15\textwidth]{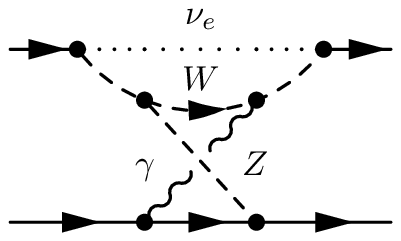}\label{FigDecoratedBoxZGUnderalW2}}
        \quad
        \subfigure[]{\includegraphics[width=0.15\textwidth]{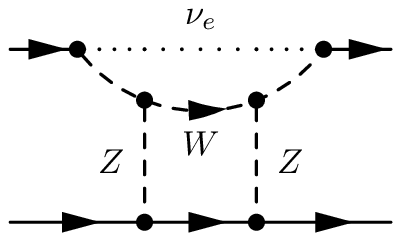}\label{FigDecoratedBoxZZUnderalW1}}
        \quad
        \subfigure[]{\includegraphics[width=0.15\textwidth]{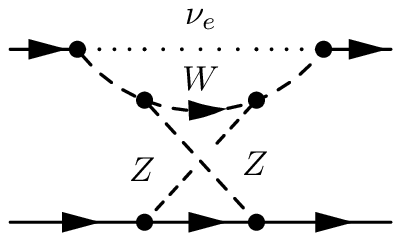}\label{FigDecoratedBoxZZUnderalW2}}
        \quad
        \subfigure[]{\includegraphics[width=0.15\textwidth]{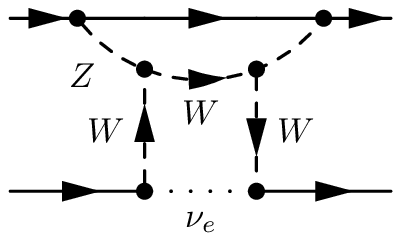}\label{FigDecoratedBoxWWUnderalZ1}}
        \quad
        \subfigure[]{\includegraphics[width=0.15\textwidth]{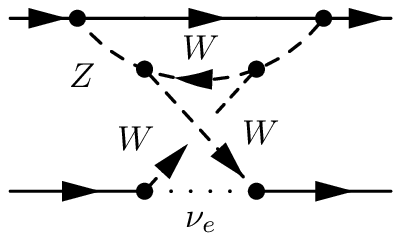}\label{FigDecoratedBoxWWUnderalZ2}}
    }
    \caption{Decorated box diagrams of Type II (see Section~\ref{SectionDecoratedBoxTypeII}).}
    \label{Fig:DecoratedBoxDiagramsTypeII}
\end{figure}
The contribution to the asymmetry from the decorated-box diagrams with two photon exchanges,
presented in Fig.~\ref{FigDecoratedBoxGGUnderalW1} and \ref{FigDecoratedBoxGGUnderalW2}, has the form:
\eq{
    A_{\gamma\gamma}^W
    =
    -\br{\frac{\alpha}{\pi}}^2  \bar A_0
    \frac{s}{4M_Z^2} \frac{1}{s_W^2}
    N_{\gamma\gamma}^W,
    \qquad
    N_{\gamma\gamma}^W
    =
    \int \frac{d\chi \, M_Z^2}{\a\,\b^2\,\a_W^2\,\c_W\,\b_e} \br{ S_1^{II}+S_2^{II} }
%    =
%    \frac{1}{c_W^2}\br{ \frac{33}{2} L_W + 30.3044 }
%    +
%    \frac{1}{c_W^2}\br{ - 2 L_W + 15\zeta_2 - 27.0073 }
    =
    \frac{1}{c_W^2}\br{ \frac{29}{2} L_W + 27.9711 },
}
with the trace $S_{1,2}^{II}$ defined in Appendix~\ref{AppendixTwoLoopsTopologiesAndTraces}.
The decorated-box diagrams with one photon and one $Z$-boson exchange,
presented in Fig.~\ref{FigDecoratedBoxGZUnderalW1}, \ref{FigDecoratedBoxGZUnderalW2},
\ref{FigDecoratedBoxZGUnderalW1} and \ref{FigDecoratedBoxZGUnderalW2},
contribute the following:
\eq{
    A_{\gamma Z}^W
    =
    - \br{\frac{\alpha}{\pi}}^2 \bar A_0
    \frac{s}{2M_Z^2} \frac{\rho c_W}{s_W^3}
    N_{\gamma Z}^W,
    \qquad
    N_{\gamma Z}^W
    =
    \int \frac{d\chi \, M_Z^2}{\a\,\b\,\b_e\,\a_W^2\,\b_Z\,\c_W} \br{ S_1^{II}+S_2^{II} }
%    =
%    38.7625 + 5.06572 + 38.7625 + 5.06572
%    = 87.6564/2
    = 43.8282.
}
The decorated-box diagrams with two $Z$-boson exchanges
(Fig.~\ref{FigDecoratedBoxZZUnderalW1} and \ref{FigDecoratedBoxZZUnderalW2})
add the term:
\eq{
    A_{ZZ}^W
    =
    - \br{\frac{\alpha}{\pi}}^2  \bar A_0
    \frac{s}{4M_Z^2} \frac{\rho^2}{s_W^4}
    N_{ZZ}^W,
    \qquad
    N_{ZZ}^W
    =
    \int \frac{d\chi \, M_Z^2}{\a\,\b_e\,\a_W^2\,\b_Z^2\,\c_W} \br{ S_1^{II}+S_2^{II} }
%    =
%    23.4951 + (-1.03141)
    = 22.4637.
}
The contribution to the asymmetry from the decorated-box diagrams with two $W$-boson exchanges (Fig.~\ref{FigDecoratedBoxWWUnderalZ1} and \ref{FigDecoratedBoxWWUnderalZ2}) is given by:
\eq{
    A_{WW}^W
    =
    - \br{\frac{\alpha}{\pi}}^2  \bar A_0
    \frac{s}{4M_Z^2} \rho^2 \frac{c_W^2}{s_W^4}
    N_{WW}^W,
    \qquad
    N_{WW}^W
    =
    \int \frac{d\chi \, M_Z^2}{\a_e\,\b\,\a_Z^2\,\b_W^2\,\c_W} \br{ S_1^{II}+S_2^{II} }
%    =
%    24.2869 + 4.61745
    = 28.9044.
}

% ==========================================================================
\subsubsection{Type III}
\label{SectionDecoratedBoxTypeIII}
% ==========================================================================
%
\begin{figure}
    \centering
    \mbox{
        \subfigure[]{\includegraphics[width=0.15\textwidth]{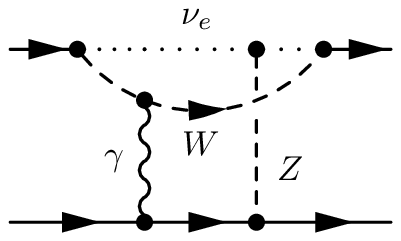}\label{FigDecoratedBoxGZBetweenW}}
        \quad
        \subfigure[]{\includegraphics[width=0.15\textwidth]{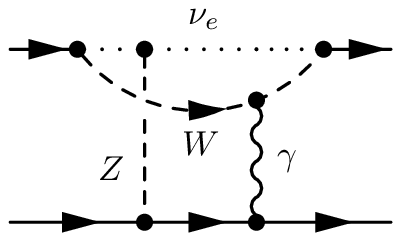}\label{FigDecoratedBoxZGBetweenW}}
        \quad
        \subfigure[]{\includegraphics[width=0.15\textwidth]{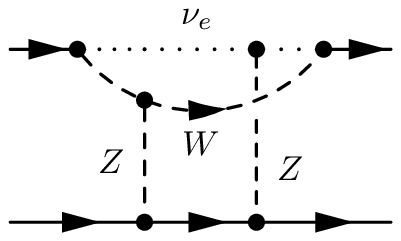}\label{FigDecoratedBoxZZBetweenW1}}
        \quad
        \subfigure[]{\includegraphics[width=0.15\textwidth]{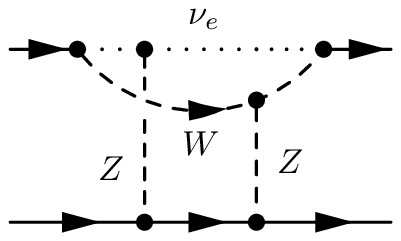}\label{FigDecoratedBoxZZBetweenW2}}
    }
    \\
    \mbox{
        \subfigure[]{\includegraphics[width=0.15\textwidth]{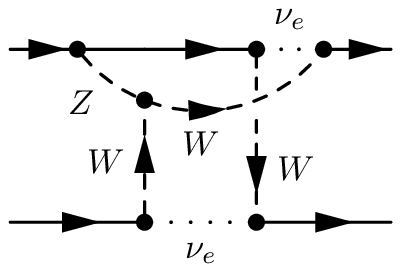}\label{FigDecoratedBoxWWBetweenW1}}
        \quad
        \subfigure[]{\includegraphics[width=0.15\textwidth]{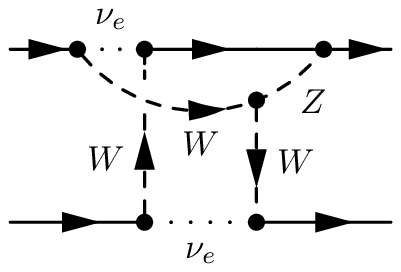}\label{FigDecoratedBoxWWBetweenW2}}
    }
    \caption{Decorated box diagrams of Type III (see Section~\ref{SectionDecoratedBoxTypeIII}).}
    \label{Fig:DecoratedBoxDiagramsTypeIII}
\end{figure}
The decorated-box diagrams with one photon and one $Z$-boson exchange,
presented in Fig.~\ref{FigDecoratedBoxGZBetweenW} and \ref{FigDecoratedBoxZGBetweenW},
give the following contribution to the asymmetry:
\eq{
    A_{\gamma Z}^W
    =
     \br{\frac{\alpha}{\pi}}^2 \bar A_0
    \frac{s}{2M_Z^2} \frac{\rho^2}{s_W^2}
    N_{\gamma Z}^W,
    \qquad
    N_{\gamma Z}^W
    =
    \int \frac{d\chi \, M_Z^2}{\a\,\b\,\c\,\c_e\,\a_W\,\b_W\,\c_Z} \br{ S_1^{III}+S_2^{III} }
%    =
%    -3.33773 + (-3.33757)
    = -6.67531,
}
with the trace $S_{1,2}^{III}$ defined in Appendix~\ref{AppendixTwoLoopsTopologiesAndTraces}.
The decorated-box diagrams with two $Z$-boson exchanges (Fig.~\ref{FigDecoratedBoxZZBetweenW1} and \ref{FigDecoratedBoxZZBetweenW2})
add the following:
\eq{
    A_{ZZ}^W
    =
    \br{\frac{\alpha}{\pi}}^2 \bar A_0
    \frac{s}{2M_Z^2} {\frac{\rho^3}{s_W^3}}
    N_{ZZ}^W,
    \qquad
    N_{ZZ}^W
    =
    \int \frac{d\chi \, M_Z^2}{\a\,\b\,\c_e\,\a_W\,\b_W\,\c_Z^2} \br{ S_1^{III}+S_2^{III} }
%    =
%    61.1774 + 61.1774
    = 122.355.
}
Finally, the contribution to the asymmetry from the decorated-box diagrams with
two $W$-boson exchanges (Fig.~\ref{FigDecoratedBoxWWBetweenW1} and \ref{FigDecoratedBoxWWBetweenW2}) has the form:
\eq{
    A_{WW}^Z
    &=
     \br{\frac{\alpha}{\pi}}^2  \bar A_0
    \frac{s}{8M_Z^2} \frac{\rho c_W}{s_W^5}
    N_{WW}^Z,
    \\
    N_{WW}^Z
    &=
    \int d\chi \, M_Z^2
    \br{
        \frac{S_1^{III}}{\a\,\b_e\,\c\,\a_W\,\b_Z\,\c_W^2}
        +
        \frac{S_2^{III}}{\a_e\,\b\,\c\,\a_Z\,\b_W\,\c_W^2}
    }
%    =
%    -1.62957 + (-1.62957)
    = -3.25915.
}
%

% =========================================================================
\section{Numerical estimation of two-loop effects to asymmetry}
\label{SectionGatheringTotalAsymmetry}
% =========================================================================

Gathering the contributions from the different Feynman diagrams considered in this paper,
we notice the following structure of the PV asymmetry $A$:
\eq{
    A
    =
    - \bar A_0 \frac{s}{M_Z^2}
    \brf{
        a_V \brs{
            R_B^Z + \frac{\alpha}{\pi} R_{\br{1}}^Z + \br{\frac{\alpha}{\pi}}^2 R_{\br{2}}^Z + \cdots
        }
        +
        \frac{\alpha}{\pi} R_{\br{1}}^W + \br{\frac{\alpha}{\pi}}^2 R_{\br{2}}^W + \cdots
    },
}
where $R_B^Z$ is the Born-level contribution to the asymmetry (\ref{AsymmetryInBorn})
and $R_{\br{1,2}}^{Z,W}$ come from the one- and two-loop radiative corrections respectively:
\eq{
    R_B^Z &= \frac{1}{2c_W^2 s_W^2} \approx \frac{8}{3} = 2.66, \nn\\
    R_{\br{1}}^Z &= 0.93775 \ln{\frac{M_Z^4}{t u}} - 0.622, \ \
    R_{\br{1}}^W = -33.4253, \label{eq:Contributions}\\
    R_{\br{2}}^Z &= -0.834656 L_Z + 40.9947, \ \
    R_{\br{2}}^W = 20.3961 L_W - 2.16291 L_Z + 99.5243. \nn
}

Translating this result to the language of the relative corrections, we
can say that the general effect of all the box contributions gives
$\delta_A = -0.93 \%$.
The main contribution to this value comes from the decorated box in
Fig. \ref{Fig:DecoratedBoxDiagramsTypeII} (a) and (b), $-0.91 \%$.
The other contributions do not exceed $\sim 0.22 \%$ and partially cancel each other.
Now, combining this value with the well-known one-loop electroweak corrections \cite{Aleksejevs:2010ub}
according to the formula (\ref{non-a}), we obtain
$\delta_A = (-0.93 \% ) \times 1.4514 = -1.35 \% $ for the central point of MOLLER kinematics.
Since the combined statistical and systematic uncertainty of MOLLER is $\delta_A^{\rm exp} \sim \pm    2 \%$,
one can clearly see that is it essential to include the two-loop radiative corrections.

% =========================================================================
\section{Conclusion}
\label{SectionConclusion}
% =========================================================================

Experimental investigation of M{\o}ller scattering is not only one of the oldest tools of modern
physics in the framework of the Standard Model, but also a powerful probe of new physics effects.
The new ultra-precise measurement of the weak mixing angle via 11 GeV M{\o}ller scattering planned
soon at JLab, named MOLLER, as well as experiments planned at the ILC, will require the higher-order
effects to be taken into account with the highest precision.

In this work,
using the on-shell renormalization scheme and the t'Hooft--Feynman gauge,
we evaluate the electroweak radiative corrections to the parity-violating asymmetry
in the \Moller scattering cross section and
account for some NNLO contributions arising from the two-loop topology of Feynman diagrams.
As one can see from our numerical data, at the MOLLER kinematic conditions the part of the
NNLO EWC we considered in this work can decrease the asymmetry by up to $\sim 1$\%.

Since the focus of this work is the two-loop box diagrams, we do not consider
the contributions arising from the SM corrections to the $W e \nu_e$ and $Wee$ vertices,
as well as SM contributions to the boson self-energies; however, they will need to be adressed in the future.
We plan to continue work in this direction and hope to provide a more precise result at some later time.

% =========================================================================
\appendix

% ==========================================================================
\section{Feynman Rules in Standard Model}
% ==========================================================================

Below we summarize a set of the SM Feynman rules relevant to our calculation of the box diagrams.
The lepton and the vector-boson propagators are:
\eq{
    \frac{i \dd{p}}{p^2 - m^2 + i0},
    \qquad
    \frac{-i g_{\mu\nu}}{k^2 - M^2 + i0},
}
where $p$ and $k$ are the lepton and boson momenta, and $m$ and $M$ are the lepton and boson masses, respectively.
We use the t'Hooft--Feynman gauge $\xi=1$.
The vector-lepton vertices are given by:
\eq{
    V(W^\mu,\nu_e, e_-)&=\frac{ie}{\sqrt{2}s_W}\gamma^\mu\omega_-; \nn \\
    V(Z^\mu, e, e)&=\frac{ie}{4s_Wc_W}\gamma^\mu(-a_V-\gamma_5); \nn \\
    V(Z^\mu,\nu_e,\nu_e)&=\frac{ie}{2s_Wc_W}\gamma^\mu\omega_-; \nn \\
    V(\gamma^\mu, e, e)&=-ie\gamma^\mu, \nn
}
where
\eq{
    a_V=1-4s_W^2, \qquad s_W=\sin\theta_W, \qquad c_W=\cos\theta_W, \qquad M_W=c_W M_Z,
    \nn
}
and $\theta_W$ is the Weinberg angle.

% ==========================================================================
\section{lepton vertex functions}
\label{AppendixVertexFunction}
% ==========================================================================

Let us calculate the correction to the electron vertex function coming from the
$Z$--boson exchange (Fig.~\ref{FigPropagatorGammaVertexZ}, \ref{FigPropagatorZVertexZ}).
Using the Feynman rules, we obtain:
\eq{
    \Gamma_\mu^Z\omega_\pm &=-ie\gamma_\mu\omega_\pm\frac{g^2(-a_V\mp 1)^2}{256\pi^2 c_W^2}I_Z, \nn \\
    I_Z&=\int\frac{d^4 k}{i\pi^2}\frac{N_Z}{(k^2-M_Z^2)(k^2-2p_1k)(k^2-2p_1'k)}, \nn \\
    N_Z&=\gamma_\lambda(\hat{p}_1'-\hat{k})\gamma_\mu(\hat{p}-\hat{k})\gamma^\lambda.
}
which leads to
\eq{
   I_Z&=
    \int\limits_0^1 dx \int\limits_0^1 2y dy
    \brf{
        \br{\ln\br{\frac{\Lambda^2}{d}}-\frac{3}{2}}
        -
        \frac{t}{2d} \br{1-y x}\br{1-y\bar x}
    },
}
where $d=y^2 p_x^2 + M_Z^2 \bar y$ and $\Lambda$ is the ultraviolet cut-off parameter.
To regularize this expression, we subtract from it the same expression but with
$t=0$.

The contribution from $W^-$ exchange, $\Gamma_\mu^W\omega_+=0$ (Fig.~\ref{FigPropagatorZVertexW}), is calculated in a similar way:
\eq{
    \Gamma_\mu^W\omega_- &=ie\frac{g^2}{32\pi^2}I_\mu^W, \nn \\
    I_\mu^W&=\int\frac{d^4 k}{i\pi^2}\frac{N_{W\mu}}{k^2(k^2-2p_1k-M_W^2)(k^2-2p_1'k-M_W^2)}, \nn \\
    N_{W\mu}&=\gamma^\nu\hat{k}\gamma^\lambda\omega_-[(p_1'-2p_1+k)^\lambda g_{\mu\nu}+
    (p_1'+p_1-2k)_\mu g_{\nu\lambda}+(p_1-2p_1'+k)_\nu g_{\mu\lambda}]. \nn
}
so we have
\eq{
    I_{W\mu}&=
    \gamma_\mu\omega_-\int\limits_0^1 dx \int\limits_0^1 2y dy
    \brs{3\br{\ln\br{\frac{\Lambda^2}{D}}-\frac{3}{2}}+\frac{t}{2D}\br{2y^2x(1-x)+y}},
}
where $D=y^2p_x^2+yM_W^2$ and $p_x^2=m^2-x(1-x)t$.

Let us consider the case of a small momentum
transfer: $-t=2p_1p_1' \ll M_W^2$. Then, the result can be summarized in the form
\eq{
    \Gamma_\mu\omega_{\pm}=-ie\gamma_\mu\omega_{\pm}\brs{1+\Delta\Gamma^Z_\pm+\Delta\Gamma^W_\pm},
}
where
\eq{
    \Delta\Gamma^W_+&=0; \nn \\
    \Delta\Gamma^W_-&=\frac{17}{18}\frac{g^2}{32\pi^2}\frac{t}{M_W^2}, \qquad -t \ll M_W^2; \nn \\
    \Delta\Gamma^Z_\pm&=\frac{g^2}{256\pi^2 c_W^2}\frac{-t}{M_W^2}(-a_V\mp 1)^2
    \br{\frac{1}{6} \ln\frac{M_Z^2}{-t}+\frac{17}{36}}. \nn
}
The contributions to the matrix element squared for a definite chiral state
\eq{
    \Delta^\lambda_i=2\M_i^\lambda(\M_B^\lambda)^*
}
are
\eq{
    \Delta^{++++}_\Gamma&=-\frac{16s^3(4\pi\alpha)^2}{t u}
    \brs{\frac{\Delta\Gamma_Z(t)}{t}+
    \frac{\Delta\Gamma_Z(u)}{u}}, \nn \\
    \Delta^{----}_\Gamma&=-\frac{16s^3(4\pi\alpha)^2}{t u}
    \brs{\frac{\Delta\Gamma_Z(t)+\Delta\Gamma_W(t)}{t}+
    \frac{\Delta\Gamma_Z(u)+\Delta\Gamma_W(u)}{u}}.
}

% ==========================================================================
\section{Contribution of $W$, $Z$ to the vertex functions of the lepton, self-energy, and vacuum polarization}
% ==========================================================================

The contributions of $WW$ and the relevant ghost intermediate states to the photon Green function
(the vacuum-polarization operator) have a form (with $\xi=1$):
\eq{
\Pi_{\mu\nu}&=-\frac{ie^2N}{32\pi^2(q^2)^2}\int\frac{d k}{(k^2-M^2)((q-k)^2-M^2)}\times\nn\\
&\times \brs{g_{\mu\nu}(2k^2-2kq+5q^2)-2q_\mu q_\nu +10k_\mu k_\nu-5(k_\mu q_\nu+k_\nu q_\mu)}+\nn\\
&+
\frac{ie^2N}{16\pi^2(q^2)^2}\int\frac{d k}{(k^2-M^2)((q-k)^2-M^2)}k_\mu(q-k)_\nu,\qquad\qquad
}
where $M=M_W$, $N=2$, $d k=d^4 k/(i\pi^2)$. Using the set of divergent integrals \cite{Akhiezer:1981},
\eq{
\int\frac{d k}{A B}&=L-1-l, \qquad A=k^2-m^2, \qquad L=\ln\frac{\Lambda^2}{m^2},  \nn \\
\int\frac{d k k_\mu}{A B}&=\frac{1}{2}\br{L-l-\frac{3}{2}}; \qquad l=\int\limits_0^1\ln(1-z x(1-x)), \qquad z=\frac{q^2}{m^2};\nn \\
\int\frac{d k k_\mu k_\nu}{A B}&=g_{\mu\nu}\brs{-\frac{1}{4}\Lambda^2+\br{-\frac{1}{12}q^2+
\frac{1}{2}m^2}L
+\br{\frac{1}{12}q^2-\frac{1}{3}m^2}l+\frac{1}{72}q^2-\frac{1}{4}m^2} + \nn \\
&+q_\mu q_\nu\brs{\frac{1}{3}L+\br{\frac{1}{3}\frac{m^2}{q^2}-\frac{1}{3}}l-\frac{5}{9}},
}
where $\Lambda$ is the ultraviolet cut-off parameter,
we obtain
\eq{
\Pi_{\mu\nu}(q)&=\frac{i\alpha}{4\pi}
\left[
    g_{\mu\nu}\brs{-4\Lambda^2+\br{-\frac{10}{3}q^2-\frac{14}{3}m^2}l+
    \br{\frac{10}{3}q^2+8m^2}L-4m^2-\frac{79}{18}q^2}+
\right.\nn  \\
&\left.
    \quad+
    q_\mu q_\nu\brs{-\frac{10}{3}L+\br{\frac{10}{3}+8\frac{m^2}{q^2}}l+\frac{32}{9}}
\right].
}
After the regularization, we have
\eq{
\Pi_{\mu\nu}(q)=\frac{i\alpha}{4\pi}\Pi\br{\frac{q^2}{m^2}}\br{q^2g_{\mu\nu}-q_\nu q_\mu}+
q_\mu q_\nu \Pi^l\br{\frac{q^2}{m^2}},
}
where
\eq{
\Pi(z)=\Pi^{SM}(z)=\br{-\frac{10}{3}+\frac{14}{3}\frac{1}{z}}\int\limits_0^1\ln(1-z x(1-x))
dx-\frac{7}{9}.
}
The term $\Pi^l(z)$ is irrelevant in our case.
Let us consider the following important limiting cases:
\eq{
\Pi^{SM}(z)&\approx -\frac{10}{3}\ln |z|+\frac{53}{9}, \qquad -z \gg 1, \nn \\
\Pi^{SM}(z)&\approx -\frac{19}{30} z+O(z^2), \qquad |z| \ll 1. \nn
}
The ultraviolet behavior is in agreement with the phenomenon of asymptotic freedom in $SU(2)$.
In our case, we use the non-relativistic limit $ m=M_{W,Z}$.
For the sake of completeness, we add the contribution of leptons calculated in the frame of QED:
\eq{
\Pi^{QED}(z)=\frac{i\alpha}{3\pi}\brs{\frac{1}{3}+\br{1+\frac{2}{z}}
\int\limits_0^1\ln(1-z x(1-x)) dx}.
}
with the limiting cases
\eq{
\Pi^{QED}(z)&=\frac{i\alpha}{3\pi}\br{\ln |z|-\frac{5}{9}}, \qquad -z \gg 1, \nn \\
\Pi^{QED}(z)&=\frac{-i\alpha}{15\pi}z, \qquad |z| \ll 1.
}

% ==========================================================================
\section{Two--loop topologies and traces}
\label{AppendixTwoLoopsTopologiesAndTraces}
% ==========================================================================

\begin{figure}
    \centering
    \mbox{
        \subfigure[]{\includegraphics[width=0.2\textwidth]{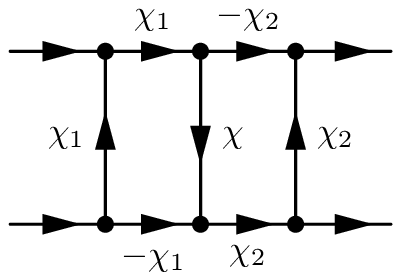}\label{Fig:Topology123}}
        \quad
        \subfigure[]{\includegraphics[width=0.2\textwidth]{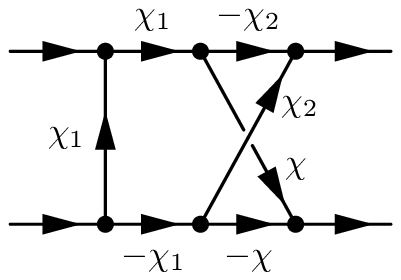}\label{Fig:Topology213}}
        \quad
        \subfigure[]{\includegraphics[width=0.2\textwidth]{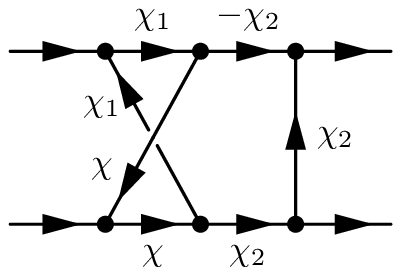}\label{Fig:Topology132}}
        \quad
        \subfigure[]{\includegraphics[width=0.2\textwidth]{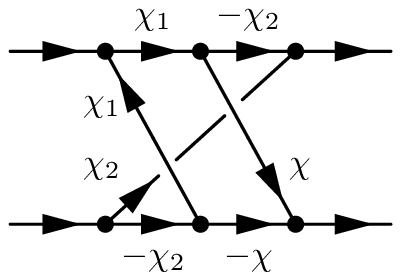}\label{Fig:Topology231}}
    }
    \\
    \mbox{
        \subfigure[]{\includegraphics[width=0.2\textwidth]{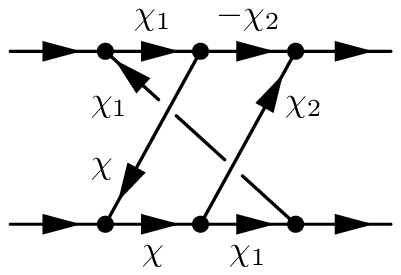}\label{Fig:Topology312}}
        \quad
        \subfigure[]{\includegraphics[width=0.2\textwidth]{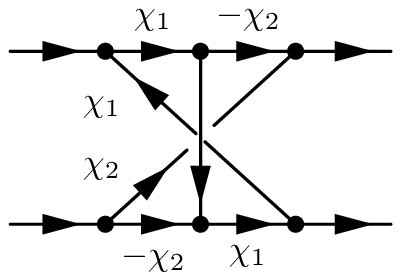}\label{Fig:Topology321}}
        \quad
        \subfigure[]{\includegraphics[width=0.2\textwidth]{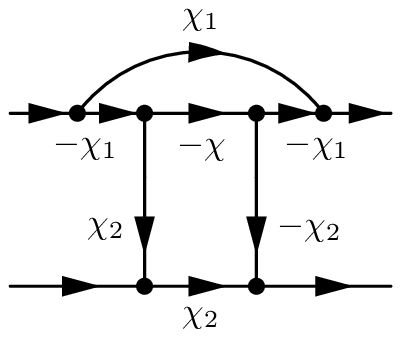}\label{Fig:TopologyDecorOveral1}}
        \quad
        \subfigure[]{\includegraphics[width=0.2\textwidth]{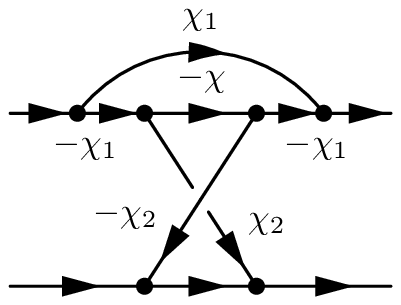}\label{Fig:TopologyDecorOveral2}}
    }
    \\
    \mbox{
        \subfigure[]{\includegraphics[width=0.2\textwidth]{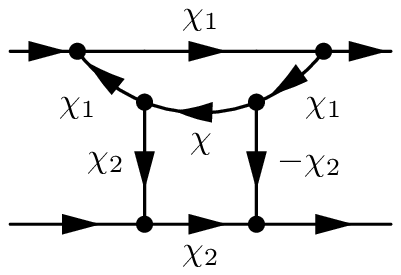}\label{Fig:TopologyDecorUnderal1}}
        \quad
        \subfigure[]{\includegraphics[width=0.2\textwidth]{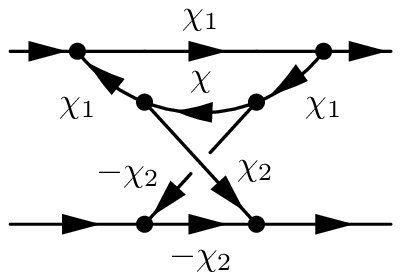}\label{Fig:TopologyDecorUnderal2}}
        \quad
        \subfigure[]{\includegraphics[width=0.2\textwidth]{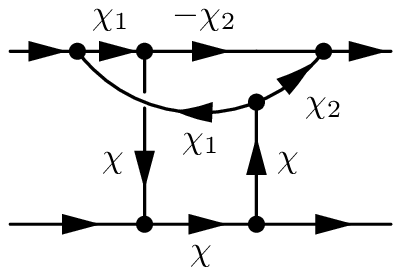}\label{Fig:TopologyDecorBetween1}}
        \quad
        \subfigure[]{\includegraphics[width=0.2\textwidth]{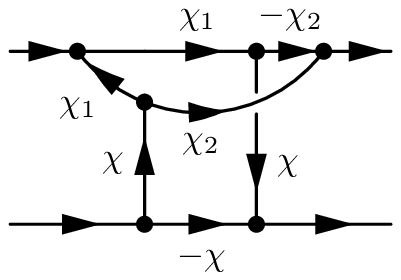}\label{Fig:TopologyDecorBetween2}}
    }
    \caption{Two-loop topology diagrams.}
    \label{Fig:TwoLoopsTopology}
\end{figure}

Here we present the traces which correspond to the topologically-different
diagrams shown in Fig.~\ref{Fig:TwoLoopsTopology}.
Let's keep in mind that the momentum integration to be done after calculating the traces neglects the dependence on the external momenta. Thus, to simplify the expressions for the
traces, we can average over momenta directions by using the following formulae:
\eq{
    \overline{\chi^\mu \chi^\nu \chi^\lambda \chi^\sigma}
    &=
    \frac{1}{24} \br{\chi^2}^2 G^{\mu \nu \lambda \sigma},
    &\qquad
    \overline{\chi_{1,2}^\mu \chi_{1,2}^\nu \chi_{1,2}^\lambda \chi_{1,2}^\sigma}
    &=
    \frac{1}{24} \br{\chi_{1,2}^2}^2 G^{\mu \nu \lambda \sigma},
    \nn\\
    \overline{\chi_1^\mu \chi_1^\nu \chi_1^\lambda \chi_2^\sigma}
    &=
    \frac{1}{24}~\alpha ~ G^{\mu \nu \lambda \sigma},
    &\qquad
    \overline{\chi^\mu \chi^\nu \chi^\lambda \chi_1^\sigma}
    &=
    \frac{1}{24}~\tilde \alpha~G^{\mu \nu \lambda \sigma},
    \nn\\
    \overline{\chi_2^\mu \chi_2^\nu \chi_2^\lambda \chi_1^\sigma}
    &=
    \frac{1}{24}~\beta~G^{\mu \nu \lambda \sigma},
    &\qquad
    \overline{\chi_1^\mu \chi_1^\nu \chi_1^\lambda \chi^\sigma}
    &=
    \frac{1}{24}~\tilde \beta~G^{\mu \nu \lambda \sigma},
    \nn\\
    \overline{\chi_1^\mu \chi_1^\nu \chi_2^\lambda \chi_2^\sigma}
    &=
    a \, g^{\mu\nu} g^{\lambda\sigma} + b \br{g^{\mu\lambda} g^{\nu\sigma} + g^{\mu\sigma} g^{\nu\lambda}},
    &\qquad
    \overline{\chi^\mu \chi^\nu \chi_1^\lambda \chi_1^\sigma}
    &=
    \tilde a \, g^{\mu\nu} g^{\lambda\sigma} + \tilde b \br{g^{\mu\lambda} g^{\nu\sigma} + g^{\mu\sigma} g^{\nu\lambda}},
    \nn
}
where $\chi=\chi_1+\chi_2$ and
\ga{
    G^{\mu \nu \lambda \sigma}
    =
    g^{\mu\nu} g^{\lambda\sigma} + g^{\mu\lambda} g^{\nu\sigma} + g^{\mu\sigma} g^{\nu\lambda},
    \nn\\
    a = \frac{1}{72} \br{5 \gamma - 2\delta},
    \qquad
    \tilde a = \frac{1}{72} \br{5 \tilde\gamma - 2\tilde\delta},
    \nn\\
    b = \frac{1}{72} \br{- \gamma + 4\delta},
    \qquad
    \tilde b = \frac{1}{72} \br{- \tilde\gamma + 4\tilde\delta},
}
and
\eq{
    \alpha &= \chi_1^2\br{\chi_1 \chi_2},
    &\quad
    \beta &= \chi_2^2\br{\chi_1 \chi_2},
    &\quad
    \gamma &= \chi_1^2\chi_2^2,
    &\quad
    \delta &= \br{\chi_1\chi_2}^2,
    \label{eq:VarsDefinition}
    \\
    \tilde\alpha &= \chi^2\br{\chi \chi_1},
    &\quad
    \tilde\beta &= \chi_1^2\br{\chi \chi_1},
    &\quad
    \tilde\gamma &= \chi^2\chi_1^2,
    &\quad
    \tilde\delta &= \br{\chi\chi_1}^2.
    \label{eq:TildeVarsDefinition}
}
The traces then read as:
\eq{
    &\text{Fig.~\ref{Fig:Topology123}}:
    &S_{123}
    &=
    \frac{1}{s^2 t}
    \Sp\brs{
        \dd{p_1'} \gamma_\mu \br{-\dd{\chi_2}} \gamma_\nu \dd{\chi_1} \gamma_\alpha \dd{p_1} \dd{p_2} \dd{p_2'}
        \gamma^\mu \dd{\chi_2} \gamma^\nu \br{-\dd{\chi_1}} \gamma^\alpha \dd{p_2} \dd{p_1} \omega_+
    }
    =32 \gamma,
    \\
    &\text{Fig.~\ref{Fig:Topology213}}:
    &S_{213}
    &=
    \frac{1}{s^2 t}
    \Sp\brs{
        \dd{p_1'} \gamma_\mu \br{-\dd{\chi_2}} \gamma_\nu \dd{\chi_1} \gamma_\alpha \dd{p_1} \dd{p_2} \dd{p_2'}
        \gamma^\nu \br{-\dd{\chi}} \gamma^\mu \br{-\dd{\chi_1}} \gamma^\alpha \dd{p_2} \dd{p_1} \omega_+
    }
    =\nn\\&&&
    =-8 (\alpha +\gamma )=8 \left(\tilde{\beta }-\tilde{\gamma }\right),
    \\
    &\text{Fig.~\ref{Fig:Topology132}}:
    &S_{132}
    &=
    \frac{1}{s^2 t}
    \Sp\brs{
        \dd{p_1'} \gamma_\mu \br{-\dd{\chi_2}} \gamma_\nu \dd{\chi_1} \gamma_\alpha \dd{p_1} \dd{p_2} \dd{p_2'}
        \gamma^\mu \dd{\chi_2} \gamma^\alpha \dd{\chi} \gamma^\nu \dd{p_2} \dd{p_1} \omega_+
    }
    =\nn\\&&&
    =-8 (\beta +\gamma )=-8 \left(\tilde{\alpha }+\tilde{\beta }-2 \tilde{\delta }\right),
    \\
    &\text{Fig.~\ref{Fig:Topology231}}:
    &S_{231}
    &=
    \frac{1}{s^2 t}
    \Sp\brs{
        \dd{p_1'} \gamma_\mu \br{-\dd{\chi_2}} \gamma_\nu \dd{\chi_1} \gamma_\alpha \dd{p_1} \dd{p_2} \dd{p_2'}
        \gamma^\nu \br{-\dd{\chi}} \gamma^\alpha \br{-\dd{\chi_2}} \gamma^\mu \dd{p_2} \dd{p_1} \omega_+
    }
    =\nn\\&&&
    =-8 (\beta +\gamma )=-8 \left(\tilde{\alpha }+\tilde{\beta }-2 \tilde{\delta }\right),
    \\
    &\text{Fig.~\ref{Fig:Topology312}}:
    &S_{312}
    &=
    \frac{1}{s^2 t}
    \Sp\brs{
        \dd{p_1'} \gamma_\mu \br{-\dd{\chi_2}} \gamma_\nu \dd{\chi_1} \gamma_\alpha \dd{p_1} \dd{p_2} \dd{p_2'}
        \gamma^\alpha \dd{\chi_1} \gamma^\mu \dd{\chi} \gamma^\nu \dd{p_2} \dd{p_1} \omega_+
    }
    =\nn\\&&&
    =-8 (\alpha +\gamma )=8 \left(\tilde{\beta }-\tilde{\gamma }\right),
    \\
    &\text{Fig.~\ref{Fig:Topology321}}:
    &S_{321}
    &=
    \frac{1}{s^2 t}
    \Sp\brs{
        \dd{p_1'} \gamma_\mu \br{-\dd{\chi_2}} \gamma_\nu \dd{\chi_1} \gamma_\alpha \dd{p_1} \dd{p_2} \dd{p_2'}
        \gamma^\alpha \dd{\chi_1} \gamma^\nu \br{-\dd{\chi_2}} \gamma^\mu \dd{p_2} \dd{p_1} \omega_+
    }
    =8 \delta,
    \\
    &\text{Fig.~\ref{Fig:TopologyDecorOveral1}}:
    &S_1^{I}
    &=
    \frac{1}{s^2 t}
    \Sp\brs{
        \dd{p_1'} \gamma_\mu \br{-\dd{\chi_1}} \gamma_\nu \br{-\dd{\chi}} \gamma_\alpha \br{-\dd{\chi_1}} \gamma^\mu \dd{p_1} \dd{p_2} \dd{p_2'}
        \gamma^\nu \dd{\chi_2} \gamma^\alpha \dd{p_2} \dd{p_1} \omega_+
    }
    =\nn\\&&&
    =-8 (\alpha +\gamma )=8 \left(\tilde{\beta }-\tilde{\gamma }\right),
    \\
    &\text{Fig.~\ref{Fig:TopologyDecorOveral2}}:
    &S_2^{I}
    &=
    \frac{1}{s^2 t}
    \Sp\brs{
        \dd{p_1'} \gamma_\mu \br{-\dd{\chi_1}} \gamma_\nu \br{-\dd{\chi}} \gamma_\alpha \br{-\dd{\chi_1}} \gamma^\mu \dd{p_1} \dd{p_2} \dd{p_2'}
        \gamma^\alpha \br{-\dd{\chi_2}} \gamma^\nu \dd{p_2} \dd{p_1} \omega_+
    }
    =\nn\\&&&
    =-4 (\alpha -\gamma +2 \delta )=4 \left(\tilde{\beta }+\tilde{\gamma }-2 \tilde{\delta }\right),
    \\
    &\text{Fig.~\ref{Fig:TopologyDecorUnderal1}}:
    &S_1^{II}
    &=
    \frac{1}{s^2 t}
    \Sp\brs{
        \dd{p_1'} \gamma_\mu \dd{\chi_1} \gamma_\nu \dd{p_1} \dd{p_2} \dd{p_2'}
        \gamma_\alpha \dd{\chi_2} \gamma_\beta \dd{p_2} \dd{p_1} \omega_+
    }
    V^{\mu\delta\alpha}\br{\chi_1,-\chi,\chi_2}
    V^{\nu\beta}_{\,\,\,\,\,\,\,\,\delta}\br{-\chi_1,-\chi_2,\chi}
    =\nn\\&&&
    =-9 (\alpha +\beta +4 \gamma ),
    \\
    &\text{Fig.~\ref{Fig:TopologyDecorUnderal1}}:
    &S_2^{II}
    &=
    \frac{1}{s^2 t}
    \Sp\brs{
        \dd{p_1'} \gamma_\mu \dd{\chi_1} \gamma_\nu \dd{p_1} \dd{p_2} \dd{p_2'}
        \gamma_\alpha \br{-\dd{\chi_2}} \gamma_\beta \dd{p_2} \dd{p_1} \omega_+
    }
    V^{\mu\delta\beta}\br{\chi_1,-\chi,\chi_2}
    V^{\nu\alpha}_{\,\,\,\,\,\,\,\,\delta}\br{-\chi_1,-\chi_2,\chi}
    =\nn\\&&&
    =15 \alpha +15 \beta +4 \gamma +20 \delta,
    \\
    &\text{Fig.~\ref{Fig:TopologyDecorBetween1}}:
    &S_1^{III}
    &=
    \frac{1}{s^2 t}
    \Sp\brs{
        \dd{p_1'} \gamma_\mu \br{-\dd{\chi_2}} \gamma_\nu \dd{\chi_1} \gamma_\alpha \dd{p_1} \dd{p_2} \dd{p_2'}
        \gamma_\beta \dd{\chi} \gamma^\nu \dd{p_2} \dd{p_1} \omega_+
    }
    V^{\mu\alpha\beta}\br{-\chi_2,-\chi_1,\chi}
    =\nn\\&&&
    =-6 (\alpha +2 \beta +\gamma +2 \delta )=-6 \left(2 \tilde{\alpha }+\tilde{\beta }-\tilde{\gamma }-2 \tilde{\delta
   }\right),
    \\
    &\text{Fig.~\ref{Fig:TopologyDecorBetween2}}:
    &S_2^{III}
    &=
    \frac{1}{s^2 t}
    \Sp\brs{
        \dd{p_1'} \gamma_\mu \br{-\dd{\chi_2}} \gamma_\nu \dd{\chi_1} \gamma_\alpha \dd{p_1} \dd{p_2} \dd{p_2'}
        \gamma^\nu \br{-\dd{\chi}} \gamma_\beta \dd{p_2} \dd{p_1} \omega_+
    }
    V^{\mu\alpha\beta}\br{-\chi_2,-\chi_1,\chi}
    =\nn\\&&&
    =-6 (2 \alpha +\beta +\gamma +2 \delta )=-6 \left(\tilde{\alpha }-\tilde{\beta }\right),
}
where the three-boson vertex $V^{\mu\nu\alpha}$ is defined for incoming momenta as the following:
\eq{
    V^{\mu\nu\alpha}\br{k_1,k_2,k_3}
    =
    \br{k_2-k_3}^\mu g^{\nu\alpha}
    +
    \br{k_3-k_1}^\nu g^{\mu\alpha}
    +
    \br{k_1-k_2}^\alpha g^{\mu\nu}.
}
We should also evaluate the trace for the diagram of $WW\gamma$ exchange
(see Fig.~\ref{Fig:LadderBoxWWGamma}):
\eq{
    S_{WW\gamma}
    &=
    \frac{1}{s^2 t}
    \Sp\brs{
        \dd{p_1'} \gamma_\mu \br{-\dd{\chi_1}} \gamma_\nu \dd{p_1} \dd{p_2} \dd{p_2'}
        \gamma_\alpha \br{-\dd{\chi}} \gamma_\beta \dd{p_2} \dd{p_1} \omega_+
    }
    V^{\mu\beta}_{\,\,\,\,\,\,\,\,\rho}\br{\chi,-\chi_1,-\chi_2}
    V^{\alpha\rho\nu}\br{-\chi,\chi_2,\chi_1}
    =\nn\\&
%    =-2 (6 \alpha -6 \beta +\gamma -4 \delta +3 \lambda )
    =2 \left(6 \tilde{\alpha }+6 \tilde{\beta }-7 \tilde{\gamma }-8 \tilde{\delta
   }\right).
   \label{eq:SpWWg}
}

% ==========================================================================
\section{Integration over the two--loop momenta}
\label{AppendixTwoLoopMomentumIntegration}
% ==========================================================================

In this section we present a set of loop integrals used to evaluate the two-loop contributions.
Although the loop momenta are normally denoted as $\chi_1$ and $\chi_2$, sometimes it is more convenient to use another set of variables,
$\chi=\chi_1+\chi_2$ and $\chi_1$.
A convenient way to classify the loop integrals is by the construction of the momenta in the numerators
(see (\ref{eq:VarsDefinition}) and (\ref{eq:TildeVarsDefinition})).
For the loop integral denominators, we use the following notation:
\eq{
    \a &= \chi_1^2,
    &\qquad
    \a_Z & = \a + M_Z^2,
    &\qquad
    \a_W & = \a + M_W^2,
    &\qquad
    \a_e = \a + m^2,
    \\
    \b &= \chi_2^2,
    &\qquad
    \b_Z & = \b + M_Z^2,
    &\qquad
    \b_W & = \b + M_W^2,
    &\qquad
    \b_e = \b + m^2,
    \\
    \c &= \chi^2,
    &\qquad
    \c_Z & = \c + M_Z^2,
    &\qquad
    \c_W & = \c + M_W^2,
    &\qquad
    \c_e = \c + m^2.
    \nn
}
Below we present the full set of master integrals used for the two-loop integration:
\eq{
    % ---------------------------------------------------
    M_Z^2 \int
    \frac{d\chi_1 d\chi_2 \, \gamma}
    {\a_e^2\,\b_e^2\,\a_Z\,\b_Z\,\c_Z}
    &=
    -
    \int\limits_0^1 dy
    \int\limits_0^1 dx
    \int\limits_0^\infty dz
    \frac{\bar x}{\br{z+1}\br{\alpha+\beta z}}
    =
    -1.87093,
    \nn\\
    % ---------------------------------------------------
    M_Z^2 \int
    \frac{d\chi_1 d\chi_2 \, \alpha}
    {\a_e^2\,\b_e\,\c_e\,\a_Z\,\b_Z\,\c_Z}
    &=
    M_Z^2 \int
    \frac{d\chi_1 d\chi_2 \, \beta}
    {\a_e\,\b_e^2\,\c_e\,\a_Z\,\b_Z\,\c_Z}
    =
    -
    \int\limits_0^1 dy
    \int\limits_0^1 dx
    \int\limits_0^\infty dz
    \frac{x^2 \bar x z}{\br{z+1}\br{a+\beta z}^2}
    =
    -0.210286,
    \nn\\
    % ---------------------------------------------------
    M_Z^2 \int
    \frac{d\chi_1 d\chi_2 \, \gamma}
    {\a_e^2\,\b_e\,\c_e\,\a_Z\,\b_Z\,\c_Z}
    &=
    M_Z^2 \int
    \frac{d\chi_1 d\chi_2 \, \gamma}
    {\a_e\,\b_e^2\,\c_e\,\a_Z\,\b_Z\,\c_Z}
    =
    -
    \int\limits_0^1 dy
    \int\limits_0^1 dx
    \int\limits_0^\infty dz
    \frac{x}{\br{z+1}\br{\alpha_y+\beta z}^2}
    =
    -1.87093,
    \nn\\
    % ---------------------------------------------------
    M_Z^2 \int
    \frac{d\chi_1 d\chi_2 \, \delta}
    {\a_e^2\,\b_e^2\,\a_Z\,\b_Z\,\c_Z}
    &=
    -
    \int\limits_0^1 dy
    \int\limits_0^1 dx
    \int\limits_0^\infty dz
    \bar y \bar x^2
    \br{
        \frac{1}{2}\frac{1}{\br{z+1}\br{\alpha+\beta z}}
        +
        x^2 \frac{z}{\br{z+1}\br{\alpha+\beta z}^2}
    }
    =
    -0.540199,
    \nn\\
    % ---------------------------------------------------
    M_Z^2 \int
    \frac{d\chi_1 d\chi_2 \, \alpha}
    {\a_e^2\,\b_e\,\c_e\,\a_Z\,\b\,\c_Z}
    &=
    \int\limits_0^1 dt
    \int\limits_0^1 dy
    \int\limits_0^1 dx
    \frac{x}{\bar x y + x t}
    =
    1.64493,
    \nn\\
    % ---------------------------------------------------
    M_Z^2 \int
    \frac{d\chi_1 d\chi_2 \, \gamma}
    {\a_e^2\,\b_e\,\c_e\,\a_Z\,\b\,\c_Z}
    &=
    -
    \int\limits_0^1 dt
    \int\limits_0^1 dy
    \int\limits_0^1 dx
    \frac{1}{\bar x y + x t}
    =
    -2\zeta_2=-3.28987,
    \nn\\
    % ---------------------------------------------------
    M_Z^2 \int
    \frac{d\chi d\chi_1 \, \tilde\alpha}
    {\a_e^2\,\b_e\,\c_e\,\a_Z\,\b_Z\,\c}
    &=
    -
    \int\limits_0^1 dt
    \int\limits_0^1 dy
    \int\limits_0^1 dx
    \frac{y}{\bar x y + x t}
    =
    -1.14493,
    \nn\\
    % ---------------------------------------------------
    M_Z^2 \int
    \frac{d\chi d\chi_1 \, \tilde\beta}
    {\a_e^2\,\b_e\,\c_e\,\a_Z\,\b_Z\,\c}
    &=
    \int\limits_0^1 dt
    \int\limits_0^1 dy
    \int\limits_0^1 dx
    \frac{x}{\bar x y + x t}
    =
    1.64493,
    \nn\\
    % ---------------------------------------------------
    M_Z^2 \int
    \frac{d\chi d\chi_1 \, \tilde\gamma}
    {\a_e^2\,\b_e\,\c_e\,\a_Z\,\b_Z\,\c}
    &=
    -
    \int\limits_0^1 dt
    \int\limits_0^1 dy
    \int\limits_0^1 dx
    \frac{1}{\bar x y + x t}
    =
    -3.28987,
    \nn\\
    % ---------------------------------------------------
    M_Z^2 \int
    \frac{d\chi_1 d\chi_2 \, \beta}
    {\a_e\,\b_e^2\,\c\,\c_e\,\a_Z\,\b_Z}
    &=
    \frac{1}{2}L_Z +
    \int\limits_0^1 dy
    \int\limits_0^1 dx
    \bar x
    \ln\br{\frac{\bar x y\br{x+\bar x y}}{\beta^2}}
    \ln\br{\frac{x+\bar x y}{\bar x y}}
    -
    \nn\\
    &-
    \int\limits_0^1 dt
    \int\limits_0^1 dy
    \int\limits_0^1 dx
    \frac{x \bar x}{\bar x y+x t}
    =
    \frac{1}{2}L_Z
    -0.289868,
    \nn\\
    % ---------------------------------------------------
    M_Z^2 \int
    \frac{d\chi_1 d\chi_2 \, \gamma}
    {\a_e\,\b_e^2\,\c\,\c_e\,\a_Z\,\b_Z}
    &=
    -L_Z -\frac{1}{2}
    \int\limits_0^1 dx
    \ln\br{\frac{x}{\beta^2}}
    \ln\br{\frac{1}{\bar x}}
    =
    -L_Z
    -2.17753,
    \nn\\
    % ---------------------------------------------------
    M_Z^2 \int
    \frac{d\chi_1 d\chi_2 \, \beta}
    {\a\,\a_e\,\b_e^2\,\c_e\,\b_Z\,\c_Z}
    &=
    \int\limits_0^1 dy
    \int\limits_0^1 dx
    \ln\br{\frac{x+y\bar x}{y\bar x}}
    =
    1.64493,
    \nn\\
    % ---------------------------------------------------
    M_Z^2 \int
    \frac{d\chi_1 d\chi_2 \, \gamma}
    {\a\,\a_e\,\b_e^2\,\c_e\,\b_Z\,\c_Z}
    &=
    -
    \int\limits_0^1 dy
    \int\limits_0^1 dx
    \int\limits_0^\infty dz
    \frac{1}{\br{z+1}\br{y + z x}}
    =
    -2\zeta_2=
    -3.28987,
    \nn\\
    % ---------------------------------------------------
    M_Z^2 \int
    \frac{d\chi_1 d\chi_2 \, \gamma}
    {\a_e^2\,\b_e^2\,\c\,\a_Z\,\b_Z}
    &=
    -
    \int\limits_0^1 dy
    \int\limits_0^1 dx
    \int\limits_0^\infty dz
    \frac{1}{\br{z+1}\br{y + z x}}
    =
    -2\zeta_2=
    -3.28987,
    \nn\\
    % ---------------------------------------------------
    M_Z^2 \int
    \frac{d\chi_1 d\chi_2 \, \delta}
    {\a_e^2\,\b_e^2\,\c\,\a_Z\,\b_Z}
    &=
    -
    \int\limits_0^1 dy
    \int\limits_0^1 dx
    \int\limits_0^\infty dz
    y\br{
        \frac{\bar x}{2}\frac{1}{\br{z+1}\br{y + z x}}
        +
        x^2\frac{z}{\br{z+1}\br{y + z x}^2}
    }
    =
    -0.572467,
    \nn\\
    % ---------------------------------------------------
    M_Z^2 \int
    \frac{d\chi d\chi_1 \, \tilde\alpha}
    {\a_e\,\b\,\c_e\,\a_W^2\,\c_W^2}
    &=
    -
    \int\limits_0^1 dy
    \int\limits_0^1 dx
    \int\limits_0^\infty dz
    \frac{y x z^2}{\br{z+1}^2\br{y + z x}^2}
    =
    -0.355066,
    \nn\\
    % ---------------------------------------------------
    M_Z^2 \int
    \frac{d\chi d\chi_1 \, \tilde\beta}
    {\a_e\,\b\,\c_e\,\a_W^2\,\c_W^2}
    &=
    -
    \int\limits_0^1 dx
    \int\limits_0^\infty dz
    \frac{x z}{\br{z+1}^2\br{1 + z x}}
    =
    -0.355066,
    \nn\\
    % ---------------------------------------------------
    M_Z^2 \int
    \frac{d\chi d\chi_1 \, \tilde\gamma}
    {\a_e\,\b\,\c_e\,\a_W^2\,\c_W^2}
    &=
    -
    \int\limits_0^1 dx
    \int\limits_0^\infty dz
    \frac{z}{\br{z+1}^2\br{1 + z x}}
    =
    -1,
    \nn\\
    % ---------------------------------------------------
    M_Z^2 \int
    \frac{d\chi d\chi_1 \, \tilde\delta}
    {\a_e\,\b\,\c_e\,\a_W^2\,\c_W^2}
    &=
    -
    \int\limits_0^1 dy
    \int\limits_0^1 dx
    \int\limits_0^\infty dz
    y \br{
        \frac{\bar x}{2}\frac{z}{\br{z+1}^2\br{y + z x}}
        +
        x^2\frac{z^2}{\br{z+1}^2\br{y + z x}^2}
    }
    =
    -0.355066,
    \nn\\
    % ---------------------------------------------------
    M_Z^2 \int
    \frac{d\chi_1 d\chi_2 \, \beta}
    {\a\,\b_e^2\,\c\,\a_W\,\b_Z\,\c_W}
    &=
    M_Z^2 \int
    \frac{d\chi_1 d\chi_2 \, \alpha}
    {\a_e^2\,\b\,\c\,\a_Z\,\b_W\,\c_W}
    =
    \int\limits_0^1 dt
    \int\limits_0^1 dy
    \int\limits_0^1 dx
    \int\limits_0^\infty dz
    \frac{x^2\bar x z}{\br{z+1}\br{a_t c_W^2 + \beta z}^2}
    =
    1.13521,
    \nn\\
    % ---------------------------------------------------
    M_Z^2 \int
    \frac{d\chi_1 d\chi_2 \, \gamma}
    {\a\,\b_e^2\,\c\,\a_W\,\b_Z\,\c_W}
    &=
    M_Z^2 \int
    \frac{d\chi_1 d\chi_2 \, \gamma}
    {\a_e^2\,\b\,\c\,\a_Z\,\b_W\,\c_W}
    =
    -
    \int\limits_0^1 dt
    \int\limits_0^1 dx
    \int\limits_0^\infty dz
    \frac{x}{\br{z+1}\br{\alpha_t c_W^2 + \beta z}}
    =
    -2.20072,
    \nn\\
    % ---------------------------------------------------
    M_Z^2 \int
    \frac{d\chi_1 d\chi_2 \, \alpha}
    {\a^2\,\b_e\,\c\,\a_W\,\b_Z\,\c_W}
    &=
    M_Z^2 \int
    \frac{d\chi_1 d\chi_2 \, \beta}
    {\a_e\,\b^2\,\c\,\a_Z\,\b_W\,\c_W}
    =
    \int\limits_0^1 dt
    \int\limits_0^1 dy
    \int\limits_0^1 dx
    \int\limits_0^\infty dz
    \frac{x^2\bar x z}{\br{z+1}\br{a_t c_W^2 + \beta z}^2}
    =
    1.13521,
    \nn\\
    % ---------------------------------------------------
    M_Z^2 \int
    \frac{d\chi_1 d\chi_2 \, \gamma}
    {\a^2\,\b_e\,\c\,\a_W\,\b_Z\,\c_W}
    &=
    M_Z^2 \int
    \frac{d\chi_1 d\chi_2 \, \gamma}
    {\a_e\,\b^2\,\c\,\a_Z\,\b_W\,\c_W}
    =
    \int\limits_0^1 dt
    \int\limits_0^1 dy
    \int\limits_0^1 dx
    \int\limits_0^\infty dz
    \frac{x\bar x z}{\br{z+1}\br{a_t c_W^2 + \beta z}^2}
    =
    2.27042,
    \nn\\
    % ---------------------------------------------------
    M_Z^2 \int
    \frac{d\chi_1 d\chi_2 \, \alpha}
    {\a^2\,\b\,\c_e\,\a_W\,\b_W\,\c_Z}
    &=
    M_Z^2 \int
    \frac{d\chi_1 d\chi_2 \, \beta}
    {\a\,\b^2\,\c_e\,\a_W\,\b_W\,\c_Z}
    =\nn\\
    &=
    \int\limits_0^1 dt
    \int\limits_0^1 dy
    \int\limits_0^1 dx
    \int\limits_0^\infty dz
    \frac{x^2\bar x z}{\br{z+c_W^2}\br{c_W^2\bar x y + x t + \beta z}^2}
    =
    1.06551,
    \nn\\
    % ---------------------------------------------------
    M_Z^2 \int
    \frac{d\chi_1 d\chi_2 \, \gamma}
    {\a^2\,\b\,\c_e\,\a_W\,\b_W\,\c_Z}
    &=
    M_Z^2 \int
    \frac{d\chi_1 d\chi_2 \, \gamma}
    {\a\,\b^2\,\c_e\,\a_W\,\b_W\,\c_Z}
    =
    \nn\\
    &=
    -
    \int\limits_0^1 dt
    \int\limits_0^1 dy
    \int\limits_0^1 dx
    \int\limits_0^\infty dz
    \frac{x \bar x z}{\br{z+c_W^2}\br{c_W^2\bar x y + x t + \beta z}^2}
    =
    -2.20072,
    \nn\\
    % ---------------------------------------------------
    M_Z^2 \int
    \frac{d\chi_1 d\chi_2 \, \delta}
    {\a^2\,\b^2\,\a_W\,\b_W\,\c_Z}
    =\nn\\
    =
    -
    \int\limits_0^1 dy
    \int\limits_0^1 dx
    \int\limits_0^\infty dz
    \bar y \bar x^2&
    \br{
        \frac{1}{2}\frac{1}{\br{z+c_W^2}\br{c_W^2 \bar x y + x + \beta z}}
        +
        x^2 \frac{z}{\br{z+c_W^2}\br{c_W^2 \bar x y + x + \beta z}^2}
    }
    =
    -0.648595,
    \nn\\
    % ---------------------------------------------------
    M_Z^2 \int
    \frac{d\chi_1 d\chi_2 \, \alpha}
    {\a_e^2\,\b^2\,\b_e\,\c\,\a_Z}
    &= \frac{1}{4} L_Z^2 + \frac{1}{4} L_Z + \zeta_2 + \frac{1}{8},
    \nn\\
    % ---------------------------------------------------
    M_Z^2 \int
    \frac{d\chi_1 d\chi_2 \, \gamma}
    {\a_e^2\,\b^2\,\b_e\,\c\,\a_Z}
    &= -\frac{1}{2} L_Z^2 - L_Z - 2\zeta_2 - \frac{1}{8},
    \nn\\
    % ---------------------------------------------------
    M_Z^2 \int
    \frac{d\chi_1 d\chi_2 \, \delta}
    {\a_e^2\,\b^2\,\b_e\,\c_e\,\a_Z}
    &= -\frac{1}{8} L_Z^2 - \frac{5}{8} L_Z - \frac{1}{2}\zeta_2 - \frac{1}{16},
    \nn\\
    % ---------------------------------------------------
    M_Z^2 \int
    \frac{d\chi_1 d\chi_2 \, \alpha}
    {\a_e^2\,\b_e\,\b\,\c_e\,\a_Z\,\b_Z}
    &=
    \int\limits_0^1 dy
    \int\limits_0^1 dx
    \int\limits_0^\infty dz
    \frac{x \bar y z}{\br{z+1}\br{y + z x}^2}
    =
    1.14493,
    \nn\\
    % ---------------------------------------------------
    M_Z^2 \int
    \frac{d\chi_1 d\chi_2 \, \gamma}
    {\a_e^2\,\b_e\,\b\,\c_e\,\a_Z\,\b_Z}
    &=
    -
    \int\limits_0^1 dy
    \int\limits_0^1 dx
    \int\limits_0^\infty dz
    \frac{1}{\br{z+1}\br{y + \bar x z}}
    =
    -3.28987,
    \nn\\
    % ---------------------------------------------------
    M_Z^2 \int
    \frac{d\chi_1 d\chi_2 \, \delta}
    {\a_e^2\,\b_e\,\b\,\c_e\,\a_Z\,\b_Z}
    &=
    -
    \int\limits_0^1 dy
    \int\limits_0^1 dx
    \int\limits_0^\infty dz
    \bar y
    \br{
        \frac{\bar x}{2} \frac{1}{\br{z+1}\br{y + x z}}
        +
        x^2 \frac{z}{\br{z+1}\br{y + x z}^2}
    }
    =
    -1.14493,
    \nn\\
    % ---------------------------------------------------
    M_Z^2 \int
    \frac{d\chi_1 d\chi_2 \, \alpha}
    {\a_e^2\,\b_e\,\c_e\,\a_Z\,\b_Z^2}
    &=
    \int\limits_0^1 dy
    \int\limits_0^1 dx
    \int\limits_0^\infty dz
    \frac{x z}{\br{z+1}^2\br{y + x z}}
    =
    0.5,
    \nn\\
    % ---------------------------------------------------
    M_Z^2 \int
    \frac{d\chi_1 d\chi_2 \, \gamma}
    {\a_e^2\,\b_e\,\c_e\,\a_Z\,\b_Z^2}
    &=
    -
    \int\limits_0^1 dy
    \int\limits_0^1 dx
    \int\limits_0^\infty dz
    \frac{z}{\br{z+1}^2\br{y + x z}}
    =
    -1.64493,
    \nn\\
    % ---------------------------------------------------
    M_Z^2 \int
    \frac{d\chi_1 d\chi_2 \, \delta}
    {\a_e^2\,\b_e\,\c_e\,\a_Z\,\b_Z^2}
    &=
    -
    \int\limits_0^1 dy
    \int\limits_0^1 dx
    \int\limits_0^\infty dz
    \bar y
    \br{
        \frac{\bar x}{2} \frac{z}{\br{z+1}^2\br{y + x z}}
        +
        x^2 \frac{z^2}{\br{z+1}^2\br{y + x z}^2}
    }
    =
    -0.572467,
    \nn\\
    % ---------------------------------------------------
    M_Z^2 \int
    \frac{d\chi_1 d\chi_2 \, \alpha}
    {\a^2\,\b_e\,\c\,\a_W\,\b_Z^2}
    &=
    \int\limits_0^1 dy
    \int\limits_0^1 dx
    \int\limits_0^\infty dz
    \frac{x z}{\br{z+1}^2\br{y c_W^2 + x z}}
    =
    0.538035,
    \nn\\
    % ---------------------------------------------------
    M_Z^2 \int
    \frac{d\chi_1 d\chi_2 \, \gamma}
    {\a^2\,\b_e\,\c\,\a_W\,\b_Z^2}
    &=
    -
    \int\limits_0^1 dy
    \int\limits_0^1 dx
    \int\limits_0^\infty dz
    \frac{z}{\br{z+1}^2\br{y c_W^2 + x z}}
    =
    -1.81942,
    \nn\\
    % ---------------------------------------------------
    M_Z^2 \int
    \frac{d\chi_1 d\chi_2 \, \delta}
    {\a^2\,\b_e\,\c\,\a_W\,\b_Z^2}
    &=
    -
    \int\limits_0^1 dy
    \int\limits_0^1 dx
    \int\limits_0^\infty dz
    \bar y
    \br{
        \frac{\bar x}{2} \frac{z}{\br{z+1}^2\br{y c_W^2 + x z}}
        +
        x^2 \frac{z^2}{\br{z+1}^2\br{y c_W^2 + x z}^2}
    }
    =
    -0.630951,
    \nn\\
    % ---------------------------------------------------
    M_Z^2 \int
    \frac{d\chi_1 d\chi_2 \, \alpha}
    {\a_e^2\,\b\,\c\,\a_Z\,\b_W^2}
    &=
    \int\limits_0^1 dy
    \int\limits_0^1 dx
    \int\limits_0^\infty dz
    \frac{1}{\br{z+c_W^2}^2\br{y + x z}}
    =
    2.36647,
    \nn\\
    % ---------------------------------------------------
    M_Z^2 \int
    \frac{d\chi_1 d\chi_2 \, \gamma}
    {\a_e^2\,\b\,\c\,\a_Z\,\b_W^2}
    &=
    -
    \int\limits_0^1 dy
    \int\limits_0^1 dx
    \int\limits_0^\infty dz
    \frac{z}{\br{z+c_W^2}^2\br{y + x z}}
    =
    -1.9256,
    \nn\\
    % ---------------------------------------------------
    M_Z^2 \int
    \frac{d\chi_1 d\chi_2 \, \delta}
    {\a_e^2\,\b\,\c\,\a_Z\,\b_W^2}
    &=
    -
    \int\limits_0^1 dy
    \int\limits_0^1 dx
    \int\limits_0^\infty dz
    \bar y
    \br{
        \frac{\bar x}{2} \frac{z}{\br{z+c_W^2}^2\br{y + x z}}
        +
        x^2 \frac{z^2}{\br{z+c_W^2}^2\br{y + x z}^2}
    }
    =
    -0.672106,
    \nn\\
    % ---------------------------------------------------
    M_Z^2 \int
    \frac{d\chi_1 d\chi_2 \, \alpha}
    {\a\,\b^2\,\b_e\,\a_W^2\,\c_W}
    &=
    \frac{1}{c_W^2}\br{\frac{1}{6}L_W + \frac{5}{18}},
    \nn\\
    % ---------------------------------------------------
    M_Z^2 \int
    \frac{d\chi_1 d\chi_2 \, \beta}
    {\a\,\b^2\,\b_e\,\a_W^2\,\c_W}
    &=
    \int\limits_0^1 dy
    \int\limits_0^1 dx
    \frac{\bar x y}{x + \bar x y}
    =
    \frac{1}{c_W^2} 0.355066,
    \nn\\
    % ---------------------------------------------------
    M_Z^2 \int
    \frac{d\chi_1 d\chi_2 \, \gamma}
    {\a\,\b^2\,\b_e\,\a_W^2\,\c_W}
    &=
    -\frac{1}{c_W^2}\br{\frac{1}{2}L_W + 1},
    \nn\\
    % ---------------------------------------------------
    M_Z^2 \int
    \frac{d\chi_1 d\chi_2 \, \delta}
    {\a\,\b^2\,\b_e\,\a_W^2\,\c_W}
    &=
    -\frac{1}{c_W^2}\br{\frac{1}{8}L_W - \frac{6}{8}\zeta_2 + \frac{13}{8}},
    \nn\\
    % ---------------------------------------------------
    M_Z^2 \int
    \frac{d\chi_1 d\chi_2 \, \alpha}
    {\a\,\b_e\,\a_W^2\,\b\,\b_Z\,\c_W}
    &=
    \int\limits_0^1 dx
    \int\limits_0^\infty dz
    \frac{x \bar x}{\br{z+1}\br{c_W^2+\beta z}}
    =
    0.405476,
    \nn\\
    % ---------------------------------------------------
    M_Z^2 \int
    \frac{d\chi_1 d\chi_2 \, \beta}
    {\a\,\b_e\,\a_W^2\,\b\,\b_Z\,\c_W}
    &=
    \int\limits_0^1 dx
    \int\limits_0^1 dy
    \int\limits_0^\infty dz
    \frac{x \bar x^2 z}{\br{z+1}\br{c_W^2\br{x+\bar x y} + \beta z}^2}
    =
    0.861218,
    \nn\\
    % ---------------------------------------------------
    M_Z^2 \int
    \frac{d\chi_1 d\chi_2 \, \gamma}
    {\a\,\b_e\,\a_W^2\,\b\,\b_Z\,\c_W}
    &=-
    \int\limits_0^1 dx
    \int\limits_0^\infty dz
    \frac{\bar x}{\br{z+1}\br{c_W^2 + \beta z}}
    =
    -1.39341,
    \nn\\
    % ---------------------------------------------------
    M_Z^2 \int
    \frac{d\chi_1 d\chi_2 \, \delta}
    {\a\,\b_e\,\a_W^2\,\b\,\b_Z\,\c_W}
    &=\nn\\
    =
    -
    \int\limits_0^1 dx
    \int\limits_0^1 dy
    \int\limits_0^\infty dz&
    y \bar x^2
    \br{
        \frac{1}{2} \frac{1}{\br{z+1}\br{c_W^2\br{x+\bar x y} + \beta z}}
        +
        x^2 \frac{z}{\br{z+1}\br{c_W^2\br{x+\bar x y} + \beta z}^2}
    }
    =
    -0.418052,
    \nn\\
    % ---------------------------------------------------
    M_Z^2 \int
    \frac{d\chi_1 d\chi_2 \, \alpha}
    {\a\,\b_e\,\a_W^2\,\b_Z^2\,\c_W}
    &=
    \int\limits_0^1 dx
    \int\limits_0^\infty dz
    \frac{x \bar x z}{\br{z+1}^2\br{c_W^2 + \beta z}}
    =
    0.250133,
    \nn\\
    % ---------------------------------------------------
    M_Z^2 \int
    \frac{d\chi_1 d\chi_2 \, \beta}
    {\a\,\b_e\,\a_W^2\,\b_Z^2\,\c_W}
    &=
    \int\limits_0^1 dx
    \int\limits_0^1 dy
    \int\limits_0^\infty dz
    \frac{y x \bar x^2  z^2}{\br{z+1}^2\br{c_W^2\br{x+\bar x y} + \beta z}^2}
    =
    0.264515,
    \nn\\
    % ---------------------------------------------------
    M_Z^2 \int
    \frac{d\chi_1 d\chi_2 \, \gamma}
    {\a\,\b_e\,\a_W^2\,\b_Z^2\,\c_W}
    &=
    -
    \int\limits_0^1 dx
    \int\limits_0^\infty dz
    \frac{\bar x  z}{\br{z+1}^2\br{1 + \beta z}}
    =
    -0.781303,
    \nn\\
    % ---------------------------------------------------
    M_Z^2 \int
    \frac{d\chi_1 d\chi_2 \, \delta}
    {\a\,\b_e\,\a_W^2\,\b_Z^2\,\c_W}
    &=\nn\\
    =
    -
    \int\limits_0^1 dx
    \int\limits_0^1 dy
    \int\limits_0^\infty dz
    y \bar x^2&
    \br{
        \frac{1}{2} \frac{z}{\br{z+1}^2\br{c_W^2 \br{x+y\bar x} + \beta z}}
        +
        x^2 \frac{z^2}{\br{z+1}^2\br{c_W^2 \br{x+y\bar x} + \beta z}^2}
    }
    =
    -0.281296,
    \nn\\
    % ---------------------------------------------------
    M_Z^2 \int
    \frac{d\chi_1 d\chi_2 \, \alpha}
    {\a_e\,\b\,\a_Z^2\,\b_W^2\,\c_W}
    &=
    \int\limits_0^1 dx
    \int\limits_0^1 dy
    \int\limits_0^\infty dz
    \frac{x \bar x^2  z^2}{\br{z+1}^2\br{c_W^2\br{x+\bar x y} + \beta z}^2}
    =
    0.674289,
    \nn\\
    % ---------------------------------------------------
    M_Z^2 \int
    \frac{d\chi_1 d\chi_2 \, \beta}
    {\a_e\,\b\,\a_Z^2\,\b_W^2\,\c_W}
    &=
    \int\limits_0^1 dx
    \int\limits_0^\infty dz
    \frac{x \bar x  z}{\br{z+1}^2\br{c_W^2 + \beta z}}
    =
    0.250133,
    \nn\\
    % ---------------------------------------------------
    M_Z^2 \int
    \frac{d\chi_1 d\chi_2 \, \gamma}
    {\a_e\,\b\,\a_Z^2\,\b_W^2\,\c_W}
    &=
    -
    \int\limits_0^1 dx
    \int\limits_0^\infty dz
    \frac{x z}{\br{z+1}^2\br{c_W^2 + \beta z}}
    =
    -0.905742,
    \nn\\
    % ---------------------------------------------------
    M_Z^2 \int
    \frac{d\chi_1 d\chi_2 \, \delta}
    {\a_e\,\b\,\a_Z^2\,\b_W^2\,\c_W}
    &=\nn\\
    =
    -
    \int\limits_0^1 dx
    \int\limits_0^1 dy
    \int\limits_0^\infty dz&
    y \bar x^2
    \br{
        \frac{1}{2} \frac{z}{\br{z+1}^2\br{c_W^2 \br{x+\bar x y} + \beta z}}
        +
        x^2 \frac{z^2}{\br{z+1}^2\br{c_W^2 \br{x+\bar x y} + \beta z}^2}
    }
    =
    -0.281296,
    \nn\\
    % ---------------------------------------------------
    M_Z^2 \int
    \frac{d\chi d\chi_1 \, \tilde \alpha}
    {\a\,\b\,\c_e\,\a_W\,\b_W\,\c_Z\,\c}
    &=
    -
    \int\limits_0^1 dx
    \int\limits_0^1 dy
    \int\limits_0^1 dt
    \int\limits_0^\infty dz
    \frac{x^2 \bar x z}{\br{z+1}\br{c_W^2 \br{\bar x y + x t} + \beta z}^2}
    =
    -1.13521,
    \nn\\
    % ---------------------------------------------------
    M_Z^2 \int
    \frac{d\chi d\chi_1 \, \tilde \beta}
    {\a\,\b\,\c_e\,\a_W\,\b_W\,\c_Z\,\c}
    &=
    -
    \int\limits_0^1 dx
    \int\limits_0^1 dy
    \int\limits_0^\infty dz
    \frac{x^2}{\br{z+1}\br{c_W^2 \br{\bar x + x y} + \beta z}}
    =
    -1.6915,
    \nn\\
    % ---------------------------------------------------
    M_Z^2 \int
    \frac{d\chi d\chi_1 \, \tilde \gamma}
    {\a\,\b\,\c_e\,\a_W\,\b_W\,\c_Z\,\c}
    &=
    -
    \int\limits_0^1 dx
    \int\limits_0^1 dy
    \int\limits_0^\infty dz
    \frac{x}{\br{z+1}\br{c_W^2 \br{\bar x + x y} + \beta z}}
    =
    -2.20072,
    \nn\\
    % ---------------------------------------------------
    M_Z^2 \int
    \frac{d\chi d\chi_1 \, \tilde \delta}
    {\a\,\b\,\c_e\,\a_W\,\b_W\,\c_Z\,\c}
    &=\nn\\
    -
    \int\limits_0^1 dx
    \int\limits_0^1 dy
    \int\limits_0^1 dt
    \int\limits_0^\infty dz&
    x\bar x
    \br{
        \frac{1}{2} \frac{1}{\br{z+1}\br{c_W^2 \br{\bar x y + x t} + \beta z}}
        +
        x^2 \frac{z}{\br{z+1}\br{c_W^2 \br{\bar x y + x t} + \beta z}^2}
    }
    =
    -1.15873,
    \nn\\
    % ---------------------------------------------------
    M_Z^2 \int
    \frac{d\chi_1 d\chi_2 \, \alpha}
    {\a\,\b\,\c_e\,\a_W\,\b_W\,\c_Z^2}
    &=
    M_Z^2 \int
    \frac{d\chi_1 d\chi_2 \, \beta}
    {\a\,\b\,\c_e\,\a_W\,\b_W\,\c_Z^2}
    =\nn\\
    &=
    -2
    \int\limits_0^1 dx
    \int\limits_0^1 dy
    \int\limits_0^1 dt
    \int\limits_0^\infty dz
    \frac{t x^3 \bar x z}{\br{z+c_W^2}\br{c_W^2 \bar x y + x t + \beta z}^3}
    =
    -0.549109,
    \nn\\
    % ---------------------------------------------------
    M_Z^2 \int
    \frac{d\chi_1 d\chi_2 \, \gamma}
    {\a\,\b\,\c_e\,\a_W\,\b_W\,\c_Z^2}
    &=
    -
    \int\limits_0^1 dt
    \int\limits_0^1 dx
    \int\limits_0^\infty dz
    \frac{t x^2 z}{\br{z+c_W^2}\br{c_W^2 \bar x + x t + \beta z}^2}
    =
    -2.61154,
    \nn\\
    % ---------------------------------------------------
    M_Z^2 \int
    \frac{d\chi_1 d\chi_2 \, \delta}
    {\a\,\b\,\c_e\,\a_W\,\b_W\,\c_Z^2}
    &=
    -2
    \int\limits_0^1 dt
    \int\limits_0^1 dx
    \int\limits_0^1 dy
    \int\limits_0^\infty dz
    t\bar x x^2
    \times\nn\\
    &\times
    \br{
        \frac{1}{4} \frac{z}{\br{z+c_W^2}\br{c_W^2 \bar x y + x t + \beta z}^2}
        +
        x^2 \frac{z^2}{\br{z+c_W^2}\br{c_W^2 \bar x y + x t + \beta z}^3}
    }
    =
    -2.96868,
    \nn\\
    % ---------------------------------------------------
    M_Z^2 \int
    \frac{d\chi d\chi_1 \, \tilde \alpha}
    {\a_e\,\b\,\c\,\a_Z\,\b_W\,\c_W^2}
    &=
    -
    \int\limits_0^1 dt
    \int\limits_0^1 dx
    \int\limits_0^1 dy
    \int\limits_0^\infty dz
    \frac{x^2 \bar x z^2}{\br{z+c_W^2}^2\br{c_W^2 x t + \bar x y + \beta z}^2}
    =
    -0.861218,
    \nn\\
    % ---------------------------------------------------
    M_Z^2 \int
    \frac{d\chi d\chi_1 \, \tilde \beta}
    {\a_e\,\b\,\c\,\a_Z\,\b_W\,\c_W^2}
    &=
    -
    \int\limits_0^1 dt
    \int\limits_0^1 dx
    \int\limits_0^\infty dz
    \frac{x^2 z}{\br{z+c_W^2}^2\br{c_W^2 x t + \bar x + \beta z}}
    =
    -1.13281,
    \nn\\
    % ---------------------------------------------------
    M_Z^2 \int
    \frac{d\chi d\chi_1 \, \tilde \alpha}
    {\a\,\b_e\,\c\,\a_W\,\b_Z\,\c_W^2}
    &=
    -
    \int\limits_0^1 dt
    \int\limits_0^1 dx
    \int\limits_0^1 dy
    \int\limits_0^\infty dz
    \frac{x^2 \bar x z^2}{\br{z+c_W^2}^2\br{c_W^2 \bar x y + t x + \beta z}^2}
    =
    -0.807308,
    \nn\\
    % ---------------------------------------------------
    M_Z^2 \int
    \frac{d\chi d\chi_1 \, \tilde \beta}
    {\a\,\b_e\,\c\,\a_W\,\b_Z\,\c_W^2}
    &=
    -
    \int\limits_0^1 dt
    \int\limits_0^1 dx
    \int\limits_0^\infty dz
    \frac{x^2 z}{\br{z+c_W^2}^2\br{c_W^2 \bar x + t x + \beta z}}
    =
    -1.06654,
    \nn\\
    % ---------------------------------------------------
    M_Z^2 \int
    \frac{d\chi d\chi_1 \, \tilde \gamma}
    {\a\,\b_e\,\c\,\a_W\,\b_Z\,\c_W^2}
    &=
    -
    \int\limits_0^1 dt
    \int\limits_0^1 dx
    \int\limits_0^\infty dz
    \frac{x z}{\br{z+c_W^2}^2\br{c_W^2 \bar x + t x + \beta z}}
    =
    -1.39341,
    \nn\\
    % ---------------------------------------------------
    M_Z^2 \int
    \frac{d\chi d\chi_1 \, \tilde \delta}
    {\a\,\b_e\,\c\,\a_W\,\b_Z\,\c_W^2}
    &=
    -
    \int\limits_0^1 dt
    \int\limits_0^1 dx
    \int\limits_0^1 dy
    \int\limits_0^\infty dz
    x\bar x
    \times\nn\\
    &\times
    \br{
        \frac{1}{2} \frac{z}{\br{z+c_W^2}^2\br{c_W^2 \bar x y + t x + \beta z}}
        +
        x^2 \frac{z^2}{\br{z+c_W^2}^2\br{c_W^2 \bar x y + t x + \beta z}^2}
    }
    =
    -0.779669,
    \nn
}
where we use the following notations:
\ga{
    a=\bar x y + x z,
    \qquad
    a_t=\bar x y + x t,
    \qquad
    \alpha=x+\bar x y,
    \qquad
    \alpha_t=\bar x + x t,
    \qquad
    \alpha_y=\bar x + x y,
    \nn\\
    \alpha_2=cx+\bar x y,
    \qquad
    \alpha_3=c_W^2 y \bar x + x,
    \qquad
    \beta = x\bar x,
    \qquad
    \bar x = 1-x. \nn
}

% ======================================================================
%\bibliographystyle{D:/Physics/Bibliography/Styles/h-physrev5}
%\bibliography{D:/Physics/Bibliography/Main}

% =========================================================================
\end{document}